\documentclass[aoas,nameyear,MSNbibl,seceqn,dvips]{arximspdf}
\usepackage{stfloats}
\usepackage{dcolumn}
\usepackage{algorithm}
\usepackage[noend]{algorithmic}
\usepackage{graphicx}

\doi{10.1214/11-AOAS457}
\volume{5}
\issue{3}
\pubyear{2011}
\firstpage{2197}
\lastpage{2230}

\makeatletter
\fnbelowfloat
\newcolumntype{d}[1]{D{.}{.}{#1}}

\newcommand{\eqref}[1]{(\ref{#1})}
\newcommand{\E}{\mathbb{E}}
\newcommand{\R}{\mathbb{R}}
\newcommand{\C}{\mathbb{C}}
\newcommand{\Prob}{\mathbb{P}}

\newtheorem{theorem}{Theorem}[section]
\newproclaim{remark}[theorem]{Remark}
\newproclaim{definition}[theorem]{Definition}
\newtheorem{corollary}[theorem]{Corollary}
\newtheorem{lemma}[theorem]{Lemma}
\newproclaim{notation}[theorem]{Notation}

\def\bsuffix #1{#1}

\def\@bmisc[#1]{%
  \get@battribute{unstr}%
  \common@pub@types%
  \let\bauthor\bbl@bauthor%
  \let\bhowpublished\@firstofone%
  \def\borganization##1{{\bauthor@style ##1}}%
}

\makeatother

\begin{document}
\begin{frontmatter}

\title{Lambert $\bolds W$ random variables---a new family of generalized skewed
distributions with applications~to risk estimation}
\runtitle{Lambert $W$ random variables}

\begin{aug}
\author[A]{\fnms{Georg M.} \snm{Goerg}\corref{}\ead[label=e1]{gmg@stat.cmu.edu}}
\runauthor{G. M. Goerg}
\affiliation{Carnegie Mellon University}
\address[A]{Department of Statistics\\
Carnegie Mellon University\\
Pittsburgh, Pennsylvania 15213\\ USA\\
\printead{e1}} 
\end{aug}

\received{\smonth{2} \syear{2010}}
\revised{\smonth{1} \syear{2011}}

%
\begin{abstract}
Originating from a system theory and an input/output point of view, I
introduce a new class of generalized distributions. A parametric
nonlinear transformation converts a random variable $X$ into a
so-called Lambert $W$ random variable $Y$, which allows a very flexible
approach to model skewed data. Its shape depends on the shape of $X$
and a skewness parameter $\gamma$. In particular, for symmetric $X$
and nonzero $\gamma$ the output $Y$ is skewed.
Its distribution and density function are particular variants of their
input counterparts. Maximum likelihood and method of moments estimators
are presented, and simulations show that in the symmetric case
additional estimation of $\gamma$ does not affect the quality of other
parameter estimates. Applications in finance and biomedicine show the
relevance of this class of distributions, which is particularly useful
for slightly skewed data. A practical by-result of the Lambert $W$
framework: data can be ``unskewed.''

The $R$ package \href{http://cran.r-project.org/web/packages/LambertW}{\texttt{LambertW}}
developed by the author is publicly available (\href{http://cran.r-project.org}{CRAN}).
\end{abstract}

%

\end{frontmatter}

\section{Introduction}
\label{sec:introduction}
Exploratory data analysis regarding asymmetry in data is usually based
on histograms and nonparametric density estimates, and statements such
as ``this data set looks almost Gaussian, but it is skewed to the
right'' or ``these asset returns have heavy tails, but they are too
skewed that a student-$t$ would make sense'' are fairly common. It is
therefore natural to generalize symmetric distributions to allow for asymmetry.

A prominent generalization is the skew-normal distribution [Azzalini\break (\citeyear{Azzalini85})],
which includes the Gaussian as a special case. A
skew-normal random variable (RV) is defined by having the probability
density function (p.d.f.) $f(x) = 2\phi(x)\Phi(\alpha x)$, where $\Phi
(\cdot)$ is the cumulative distribution function (c.d.f.) of a standard
Gaussian, and $\alpha\in\R$ is the skew parameter. This approach to
skewness has not only led to substantial research in the skew-normal
case [\citet{AzzaliniCapitanio99}, \citet{ArellanoAzzalini06}], but the same
concept has also been used for student-$t$ [\citet{AzzaliniCapitanio03}]
and Cauchy distributions [\citet{ArnoldBeaver00},\vadjust{\goodbreak} \citet{BehboodianJamalizadehBalakrishnan06}], among others. In all these
cases, a parametric manipulation of the original symmetric p.d.f.
introduces skewness.

Notwithstanding the huge success of this approach to model skewed data,
manipulating the p.d.f. to introduce skewness seems like putting the cart
before the horse: densities are skewed, because the random variable is---not the other way around. Also, applied research starts with data,
not with histograms.

\begin{figure}

\includegraphics{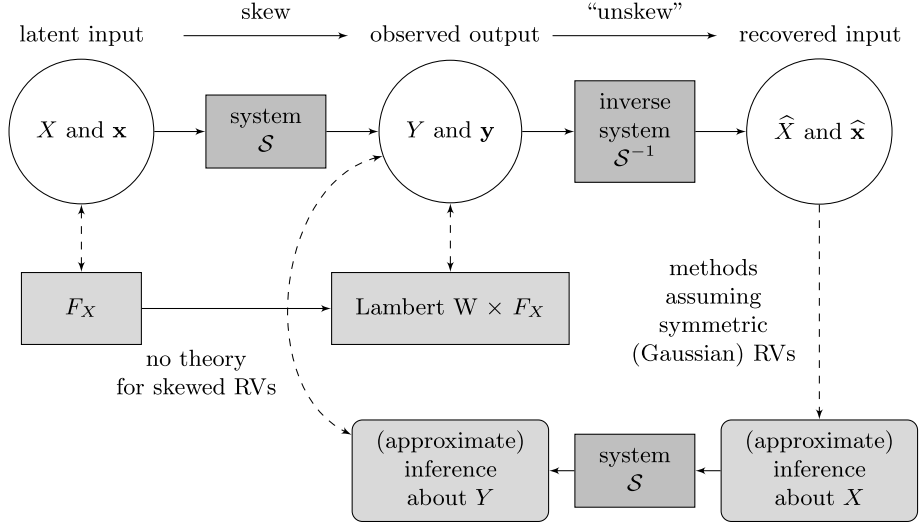}

\vspace*{-2pt}
\caption{Schematic view of the Lambert
$W$ approach to asymmetry: (left) an input/output system $\mathcal{S}$
transforms (solid arrows) input $X \sim F_X$ to output $Y \sim$
Lambert $W\times F_X$ and herewith introduces skewness; (right)
inference about skewed data $\mathbf{y} = (y_1, \ldots , y_N)$:
(1)~unskew observed $\mathbf{y}$ to latent symmetric data $\widehat
{\mathbf{x}}$, (2) use methods of your choice (regression, time series
models, quantile estimation, hypothesis tests, etc.) for statistical
inference based on $\widehat{\mathbf{x}}$, and (3) convert results
back to the ``skewed world'' of $\mathbf{y}$.}\label{fig:LambertW_flowchart}
\vspace*{-2pt}
\end{figure}

Motivated by this data-driven view on skewness, I propose a novel
approach to asymmetry: Lambert $W \times F_X$ distributions emerge
naturally by modeling the observable RV $Y$ as the output of a system
$\mathcal{S}$ driven by random input $X$ with c.d.f. $F_X(x)$. Here
$\mathcal{S}$ can either be a real chemical, physical, or biological
system, or refer to any kind of mechanism in a broader sense. In
statistical modeling such a system is simply represented by
transformations of RVs. As there are no restrictions on $F_X(x)$, this
is a very general framework that can be analyzed in detail for a
particularly chosen input c.d.f. Figure \ref{fig:LambertW_flowchart}
illustrates the methodology.

\begin{figure}[b]
\vspace*{-2pt}

\includegraphics{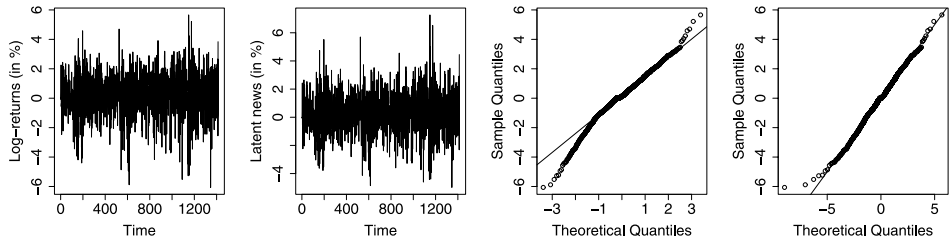}
\vspace*{-2pt}

\caption{Daily equity fund returns (LATAM):
({left}) observed returns $\mathbf{y}$ and estimated latent news
$\widehat{\mathbf{x}}_{\widehat{\tau}_{\mathrm{MLE}}}$; ({right})
Gaussian QQ plot and Lambert $W\times t$ QQ plot of $\mathbf
{y}$.}\label{fig:LATAM_news4}
\end{figure}

For instance, consider $\mathcal{S}$ being the stock market, where
people buy and sell an asset according to its expected success in the
future. Asset returns, that is, the percentage change in price,
typically exhibit negative skewness and positive excess kurtosis---so-called
\textit{stylized facts} [\citet{Yan05},
\citet{Cont01}]. The left
panel of Figure \ref{fig:LATAM_news4} shows daily log-returns (in
percent) $\mathbf{y} := \lbrace y_t \mid t=1, \ldots , 1\mbox{,}413 \rbrace$
of an equity fund investing in Latin America (LATAM\footnote{Data from
January 1, 2002 until May 31, 2007: R package \texttt{fEcofin}, data
set \texttt{equityFunds}, series \texttt{LATAM}.}). Also, these
returns are clearly non-Gaussian given their excess kurtosis ($1.201$)
and large negative skewness ($-0.433$)---see Table \ref
{tab:LATAM_summary}. The excess kurtosis is typically addressed by a
student $t$-distribution, but here a Kolmogorov--Smirnov test still
rejects $\mathbf{y} \sim t_{6.22}$ on a $5\%$ level (even for the
estimated~$\nu$), as the empirical skewness is too large. Thus, to
model the probabilistic properties of such data, asymmetric
distributions must be used.

%
%

Using Lambert $W\times F$ RVs to model the asymmetry in asset returns
is perfectly suitable not only given empirical evidence of ``almost
student-$t$, but a little skewed data,'' but also by a more fundamental
viewpoint. Price changes are commonly considered as the result of bad
and good news hitting the market: bad news, negative returns; good
news, positive returns. The empirical evidence of negative skewness
evokes the following question: why should news per se be
negatively skewed? Or put in other words: do really bad things happen
more often than really good things?

In the Lambert $W$ framework this news $\leftrightarrow$ return relation
is modeled under the assumption that the probability of getting
negative news is about the same as of getting positive news, but
typically people react far more drastically facing negative than
positive ones. Thus, news $X \sim F_X(x)$ are symmetrically
distributed, the market $\mathcal{S}$ acts as an asymmetric filter,
and the measurable/observable outcome is a skewed RV $Y$/data $\mathbf
{y}$.\vadjust{\eject}

Last, the right part of Figure \ref{fig:LambertW_flowchart} also
illustrates a very pragmatic, yet useful way to exploit the Lambert $W$
framework for (slightly) skewed data. If a~certain statistical
procedure or model assumes a symmetric---a Gaussian, as often is the
case---distribution and no skewed implementation of this method is
available, then instead of applying it to the skewed $\mathbf{y}$, it
is advisable to work with the ``symmetrized'' data $\widehat{\mathbf
{x}}$, make statistical inference about $X$ based on $\widehat{\mathbf
{x}}$, and then transform the obtained results back to the ``skewed
world'' of $Y$. Although this is only an approximation to the truth, at
least this approach takes skewness into consideration instead of~ignoring~it.

%
\begin{table}
\tabcolsep=0pt
\tablewidth=275pt
\caption{LATAM returns $\mathbf{y}$ and
back-transformed series $\widehat{\mathbf{x}}_{\widehat{\tau}}$:
(top)~summary~statistics; (bottom) Shapiro--Wilk (SW) $\&$ Jarque--Bera
(JB) normality tests, and Kolmogorov--Smirnov (KS) test for student-$t$
with~$\nu$~degrees~of~freedom}\label{tab:LATAM_summary}
\begin{tabular*}{275pt}{@{\extracolsep{\fill}}ld{2.3}d{2.3}d{2.3}@{}}
\hline
 \textbf{LATAM}  & \multicolumn{1}{c}{$\mathbf{y}$} &
\multicolumn{1}{c}{$\bolds{\widehat{\mathbf{x}}_{{\widehat\tau_{\mathrm
{IGMM}}}}}$} & \multicolumn{1}{c@{}}{$\bolds{\widehat{\mathbf{x}}_{
{\widehat\tau_{\mathrm{MLE}}}}}$} \\
\hline
Min & -6.064 & -5.073 & -4.985 \\
Max & 5.660 & 7.036 & 7.268 \\
Mean & 0.121 & 0.190 & 0.198 \\
Median & 0.138 & 0.138 & 0.138 \\
St. dev. & 1.468 & 1.456 & 1.457 \\
Skewness & -0.433 & 0.000 & 0.053 \\
Kurtosis & 1.201 & 1.100 & 1.159 \\
$\widehat{\nu}$ & 6.220 & 7.093 & 7.048 \\
 [6pt]
SW & 0.000 & 0.000 & 0.000 \\
JB & 0.000 & 0.000 & 0.000 \\
 [3pt]
KS ($t_{\widehat{\nu}}$) & 0.028 & 0.088 & 0.102\\
\hline
\end{tabular*}
\end{table}

Section \ref{sec:LambertW_RV} defines Lambert $W$ RVs and their basic
properties are studied. Section~\ref{sec:dist_dens} presents analytic
expressions of the c.d.f. $G_Y(y)$ and p.d.f. $g_Y(y)$, which are particular
variants of their input counterparts. After studying Gaussian input in
Section \ref{sec:Gaussian}, various estimators for the parameter
vector of Lambert $W\times F$ RVs are introduced in Section \ref
{sec:estimation}. 
Section \ref{sec:simulations} compares their finite sample properties
and shows that additional estimation of the skewness parameter $\gamma
$ does not affect the quality of other parameter estimates. This new
class of distribution functions is particularly useful for data with
slightly negative skewness, thus, Section \ref{sec:applications}
demonstrates its adequacy on an Australian athletes data set and on the
LATAM return series.

In particular, Section \ref{sec:financial_data} shows that the
input-output system (Figure~\ref{fig:LambertW_flowchart}) with
student-$t$ input $X$ is a proper model for these returns. A detailed
comparison of quantile estimates, which are essential to get
appropriate risk measures of an asset, confirms the aptness of Lambert
$W$ distributions (see Lambert $W$ QQ plot in Figure \ref
{fig:LATAM_news4}). Empirical evidence for the significance of
conditional heteroskedastic time series models using Lambert $W\times
F$ innovations concludes Section \ref{sec:financial_data}.

Finally, Section \ref{sec:Tukeys_h} establishes a direct link of this
new class of distributions to the existing statistics literature,
noting that the square of a RV having Tukey's $h$ distribution [\citet
{Tukey77}] has a Lambert $W \times \chi_1^2$ distribution.

Computations, figures and simulations were realized with the
open-source statistics package R [\citet{R08}]. Functions used in the
analysis are available as the R package
\href{http://cran.r-project.org/web/packages/LambertW/index.html}{\texttt{LambertW}}, which provides many other methods to perform Lambert $W$
inference in practice.

\vspace*{5pt}
\section{Lambert $W$ random variables}
\label{sec:LambertW_RV}

The general notion of a system $\mathcal{S}$ with random input and
output as shown in Figure \ref{fig:LambertW_flowchart} translates to
a variable transformation in statistical terminology.

\begin{definition}[(Noncentral, nonscaled Lambert $W\times F$ RV)]
\label{def:noncentral_nonscaled_LambertW}
Let $U$ be a continuous RV with c.d.f.
%
%
\begin{equation}
\label{eq:cdf_of_U}
F_U(u) = \Prob(U \leq u), \qquad u \in\R,
\end{equation}
and p.d.f. $f_U(u)$. Then
%
%
\begin{equation}
\label{eq:noncentral_nonscale_LambertW_Y}
Z := U \exp(\gamma U), \qquad\gamma\in\R,
\end{equation}
is a \textit{noncentral, nonscaled Lambert $W\times F$} RV with
skewness parameter $\gamma$.

If $U$ is from a parametric family $F_U(u \mid\bolds\beta)$,
where $\bolds\beta$ parametrizes the $F_U$, then $Z$ is a
noncentral, nonscaled Lambert $W\times F$ RV with parameter vector
$\theta= (\bolds\beta, \gamma)$.
\end{definition}

The key of this family of RVs is $\gamma$, which can take any value on
the real line. As $\exp(\cdot)$ is always positive, $U$ and $Z$ have
the same sign. For readability let $H_{\gamma}(u) := u \exp(\gamma
u)$. For $\gamma= 0$ transformation \eqref
{eq:noncentral_nonscale_LambertW_Y} reduces to the identity $Z \equiv
U$; thus, $Z$ possesses the exact same properties as $U$. By continuity
of $H_{\gamma}(\cdot)$, one can expect for $\gamma\neq0$ but close,
also $Z \neq U$ but close.\looseness=1

Transformation \eqref{eq:noncentral_nonscale_LambertW_Y} indeed
describes a system $\mathcal{S}$ with an asymmetry property: let $U
\sim F_U(u)$ be a symmetric zero-mean RV, then $Z$ is a skewed version
of $U$---depending on the sign of $\gamma$. For $\gamma<0$ negative
$U$ are amplified by the factor $\exp(\gamma U) > 1$ and positive $U$
are damped by \mbox{$0<\exp(\gamma U) <1$}: $Z$ is skewed to the left. For
$\gamma>0$ the same reasoning shows that $Z$ is a~positively.

The noncentral moments $\E ( Z^n  ) $ equal
%
%
\begin{equation}
\label{eq:non_contral_mom_Z}
\psi_{(n)} 
:= \int u^n e^{\gamma u n} f_U(u) \,\mathrm{d} u.
\end{equation}

If the moment-generating function $M_U(t):= \E e^{t U}$ exists for $t =
\gamma n$, then \eqref{eq:non_contral_mom_Z} can be rewritten to get a
more tractable formula. As
\[
\frac{\partial^n}{\partial\gamma^n} e^{\gamma U n} = (U n)^n
e^{\gamma U n},
\]
interchanging differentiation and the integral sign yields
%
%
\begin{eqnarray}
\label{eq:non_central_mom_Z_derivative}
\psi_{(n)}
&=& n^{-n} \,\frac{\partial^n}{\partial\gamma^n} M_U(\gamma n).
\end{eqnarray}

If $M_U(t)$ does not exist (e.g., for student-$t$ $U$), then
\eqref{eq:non_contral_mom_Z} must be calculated explicitly.

\vspace*{-4pt}
\subsection{Scale family input}
In a typical input/output system $\mathcal{S}$ such as
a~microphone/loudspeaker setting, the loudspeaker will be louder if
speakers raise their voice. In this sense it is stable with respect to
scaling: doubling the volume of the input doubles the volume of the
loudspeakers---the signal is not affected in any other way. Viewing
this system as a Lambert $W\times F$ RV system (where the signal is
considered as a RV), multiplying $X$ by a~factor~$\kappa$,
should---ceteris paribus---only affect the output $Y$ by multiplying by~$\kappa$;
other properties, such as skewness or kurtosis, should not be
altered.\looseness=-1

Transformation \eqref{eq:noncentral_nonscale_LambertW_Y}, however,
does not have this scaling property of $U$. Hence, to allow a
comparable system characterization via $\gamma$ among different
scalable data sets define a scaled Lambert $W$ RV.

\vspace*{-4pt}
\begin{definition}[(Scale Lambert $W\times F$ RV)]
\label{def:scale_LambertW}
Let $U := X / \sigma_x$ be the unit-variance version of a continuous
RV $X$ from a scale family $F_X  ( x \mid\bolds\beta
)$, where $\bolds\beta$ is the parameter (vector) of $F_X$ and
$\sigma_x$ the standard deviation of~$X$. Then
%
%
\begin{equation}
\label{eq:scaled_LambertW_Y}
Y :=  \{ U \exp(\gamma U )  \} \sigma_x = X \exp(\gamma X
/ \sigma_x), \qquad\gamma\in\R,\ \sigma_x >0,
\end{equation}
is a \textit{scale Lambert $W\times F$} RV with parameter vector
$\theta= (\bolds\beta, \gamma)$.
\vspace*{-4pt}
\end{definition}

Transformation \eqref{eq:scaled_LambertW_Y} is invariant to scaling of
the input, for example, a~different measurement unit for the input
does not modify the asymmetry property of the system, but just scales
the output accordingly.

Here $\sigma_x$ is a function of $\bolds\beta$: for an
exponentially distributed input $X \sim\exp(\lambda)$, $\bolds
\beta= \lambda$ and $\sigma_x(\bolds\beta) = \lambda^{-1}$;
an input $X$ having a Gamma distribution with shape $\alpha$ and rate
$\beta$ gives $\bolds\beta= (\alpha, \beta)$ and $\sigma
_x(\bolds\beta) = \sqrt{\alpha} / \beta$.

\vspace*{-4pt}
\subsection{Location-scale family input}
The focus of this work lies in introducing skewness to symmetric RVs
with support on $(-\infty, \infty)$, such as a~Gaussian or student-$t$.
These distributions are not only scale, but also shift invariant, a
property Lambert $W\times F$ distribution should also have for
location-family input. However, transformation \eqref
{eq:scaled_LambertW_Y} is not shift-invariant. For example, consider a
zero-mean and unit variance input RV $U_0\dvtx  \Omega\rightarrow\R$,
$U_{10} := U_0 + 10$, and let $\gamma=0.1$. If $U_0(\omega)$ is close\vadjust{\eject}
to $0$, then the shifted $U_{10}(\omega)$ will be close to $10$. For
the corresponding $Z_{0}(\omega)$ and $Z_{10}(\omega)$ this does not
hold: $Z_0(\omega)$ is close to $0$, but $Z_{10}(\omega)$ will not be
shifted by $10$, but lies close to $10 \exp(1) \approx27.183$.

\vspace*{-4pt}
\begin{definition}[(Location-scale Lambert $W\times F$ RV)]
\label{def:location_scale_LambertW}
Let $X$ be a RV from a location-scale family with c.d.f. $F_X ( x
\mid\bolds\beta )$ with mean $\mu_x$ and standard
deviation $\sigma_x$; again $\bolds\beta$ parametrizes $F_X$.
Let $U = (X - \mu_x)/ \sigma_x$ be the zero-mean, unit-variance
version of $X$. Then
%
%
\begin{equation}
\label{eq:location_scale_LambertW_Y}
Y :=  \{ U \exp(\gamma U )  \} \sigma_x + \mu_x, \qquad
\gamma\in\R,
\end{equation}
is a \textit{location-scale Lambert $W\times F$} RV with parameter
vector $\theta= (\bolds\beta, \gamma)$.
\vspace*{-4pt}
\end{definition}

As before, the parameter $\gamma$ regulates the closeness between $X$
and its skewed version $Y$.

For a full parametrization of a Lambert $W\times F$ distribution it
is necessary to know $\theta= (\bolds\beta, \gamma)$; viewing
\eqref{eq:location_scale_LambertW_Y} only as a transformation from $X$
to $Y$, it is more natural---and in practice more useful---to only
consider $\mu_x$, $\sigma_x$ and~$\gamma$, ignoring the particular
structure of $X$ given its parametrization by~$\bolds\beta$. In
order to distinguish these two cases in the remaining part of this work
let $\tau:= (\mu_x, \sigma_x, \gamma) \in T = \R\times\R^{+}
\times\R$. Clearly, $\tau$ can be computed from $\theta$, $\tau=
\tau(\theta)$, but not necessarily vice-versa.

For example, for a Gaussian $X$ $\tau\equiv\theta$ since $\mu
_x(\bolds\beta) = \mu_x$ and $\sigma_x(\bolds\beta) =
\sigma_x$. In contrast, for a location-scale student-$t$ input with
$\bolds\beta= (c, s, \nu)$---where $c$ is the location, $s$ the
scale and $\nu$ the degrees of freedom parameter---$\tau\neq\theta
$: $\mu_x(\bolds\beta) = c$ and $\sigma_x(\bolds\beta)
= s \sqrt{\frac{\nu}{\nu-2}}$ if $\nu> 2$.

Thus, below I use either $\theta$ if the full parametrization is
important or $\tau$ if it is sufficient to consider \eqref
{eq:location_scale_LambertW_Y} only as a transformation rather than a
fully specified parametric distribution.

\vspace*{-4pt}
\begin{notation}[(Lambert $W\times F$ RV)]
\label{notation:LambertW_F}
For simplicity I will refer to all~$Y$ in Definitions \ref
{def:noncentral_nonscaled_LambertW}, \ref{def:scale_LambertW} and \ref
{def:location_scale_LambertW} as a \textit{Lambert $W\times F$} RV.
Which one of the three transformations \eqref
{eq:noncentral_nonscale_LambertW_Y}, \eqref{eq:scaled_LambertW_Y} or
\eqref{eq:location_scale_LambertW_Y} is used to generate $Y$ will be
clear from the type of input $X$. For example, since a
$\chi_k^2$\vspace*{1pt}
distribution does not have location or scale parameters, a Lambert $W\times\chi_k^2$ RV refers to $Y$ in Definition \ref
{def:noncentral_nonscaled_LambertW}; the exponential distribution is a
scale family, thus, a Lambert $W\times\exp(\lambda)$ RV $Y$ is
defined in Definition~\ref{def:scale_LambertW}; and for Gaussian input $X$, the
corresponding Lambert $W{}\times{}$Gaussian $Y$ refers to Definition
\ref{def:location_scale_LambertW}.\footnote{Although technically not
correct, one can think of a scale Lambert $W\times F$ transformation
having $\tau= (0, \sigma_x, \gamma)$, and a noncentral, nonscaled
Lambert $W\times F$ transformation having $\tau= (0, 1, \gamma)$.
This is especially useful for empirical work and implementation of the
methods involving Lambert $W\times F$ RVs.}
\vspace*{-4pt}
\end{notation}

%
\subsection{Latent variables}
So far attention has been drawn to $Y$ and its properties given $X$ and
$\theta$ (or $\tau$). Now consider the inverse problem: given $Y$ and~$\theta$ (or only~$\tau$),
what does $X$ look like?\vadjust{\eject}

\begin{figure}

\includegraphics{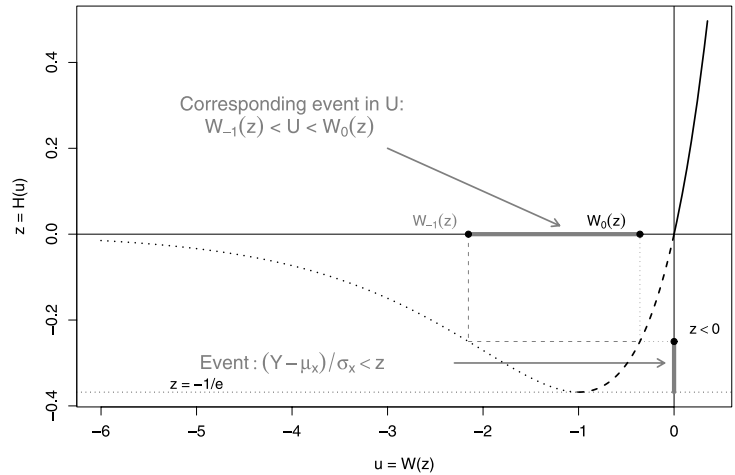}

\caption{Lambert $W$ function:
transformation $H(u)$ and the two inverse branches of $W(z)$ for $z<0$:
principal branch (dashed curve) and nonprincipal branch (dotted
curve).}\label{fig:Events_Z_Events_U}
\vspace*{-4pt}
\end{figure}

This is not only interesting for a latent variable interpretation of
$X$, but the inverse of a transformation is essential to derive the c.d.f.
of the transformed variable. Before analyzing transformation \eqref
{eq:location_scale_LambertW_Y}, consider the nonlinear transformation
$H\dvtx  \C\rightarrow\C, u \mapsto u \exp{u} =:z$ [Figure \ref
{fig:Events_Z_Events_U} shows $H(u)$ only for $u \in\R$]. For
positive $u$ the function is bijective and resembles $\exp(u)$ very
closely. For negative $u$, however, $H(u)$ is quite different from
$\exp(u)$: it takes on negative values, its minimum value equals
$-\frac{1}{e}$, and---most importantly---it is nonbijective.

%

Although $H(u)$ has no analytical inverse [\citet{Rosenlicht69}], its
implicitly defined inverse function is well known in mathematics and physics.
\begin{definition}[(Lambert $W$ function)]
The many-valued function $W(z)$ is the root of
%
%
\begin{equation}
\label{eq:W_implicit_sol}
W(z) e^{W(z)} = z, \qquad z \in\C,
\end{equation}
and is commonly denoted as the \textit{Lambert $W$ function}.
\end{definition}

Generally the Lambert $W$ function is defined for any $z \in\C$. Since
Lambert $W$ RVs are only defined for real-valued outcomes, in this work
the domain and image of the Lambert $W$ function is restricted to the
reals. For $z \in[-\infty,-1/e)$ no real solution exists; for $z \in
[-1/e, \infty)$ $W(z)$ is a real-valued function. If $z \in[-1/e,0)$,
there are two real solutions: the principal branch $W_0(z) \geq-1$ and
the nonprincipal branch $W_{-1}(z) \leq-1$; for $z \in[0,\infty)$
only one real-valued solution exists, $W_0(z) = W_{-1}(z)$ (see Figure
\ref{fig:Events_Z_Events_U}). 

For a detailed review including useful properties and functional
identities of $W(z)$ see \citet{Corless96}, \citet{ValluriJeffreyCorless00}
and the references therein.\vadjust{\eject}

Figure \ref{fig:Events_Z_Events_U} also shows how skewness is
introduced via transformation \eqref{eq:location_scale_LambertW_Y}.
Symmetric input $X$ ($x$-axis) is mapped to asymmetric output $Y$
($y$-axis) due to the curvature of $H(u)$. Analogously, mapping values
from the $y$-axis to the $x$-axis ``unskews'' them. Figure \ref
{fig:Events_Z_Events_U} shows $H_{\gamma}(u)\dvtx  u \mapsto z$ for $\gamma
= 1$, thus, its inverse is Lambert's $W$ function ($W(z)\dvtx  z \mapsto u$).
The curvature of $H_{\gamma}(u) = u \exp(\gamma u)$ depends on the
skewness parameter: for $\gamma= 0$ no curvature is present [$H_0(u) =
u$]; higher $\gamma$ results in more curvature, and thus more skewness
in $Y$.

It can be easily verified that $W_{\gamma}(Z) : = W(\gamma Z) / \gamma
$ is the inverse function of transformation \eqref
{eq:noncentral_nonscale_LambertW_Y}. Hence, given $Y$ and $\tau$,
the unobservable input $X$ can be recovered via
%
%
\begin{equation}
\label{eq:backtrafo_normalized_X}
W_{\gamma} \biggl(\frac{Y - \mu_x}{\sigma_x} \biggr) \sigma_x + \mu
_x = U \sigma_x + \mu_x = X.
\end{equation}
For empirical work it is important to point out that \eqref
{eq:backtrafo_normalized_X} does not require specific knowledge about
$F_X$ or $\bolds\beta$; $\mu_x$ and $\sigma_x$ (and $\gamma$)
suffice. This will become especially useful for estimating the optimal
inverse transformation---see Section \ref{sec:IGMM}.

\begin{remark}[(Nonprincipal branch)]
The Lambert $W$ function has two branches on the negative real line
(Figure \ref{fig:Events_Z_Events_U}), so transformation \eqref
{eq:backtrafo_normalized_X} is not unique. For example, consider $z =
-0.25$ and $\gamma= 1$. The two real-valued solutions are $W_0(-0.25)
\approx-0.357$ and $W_{-1}(-0.25) \approx-2.153$. Assuming a stable
input/output system, only the principal branch makes sense\footnote
{The output is assumed to be similar to the input, but skewed.
Therefore, the input values causing the output should lie close to
them: observing $z = -0.25$, it is more reasonable to assume that this
corresponds to the close input of $W_0(-0.25) \approx-0.357$ rather
than the very extreme $W_{-1}(-0.25) \approx-2.153$.}---denoted by
$W_{\gamma, 0}(\cdot)$. If the nonprincipal solution is required,
$W_{\gamma, -1}(\cdot)$ will be used.

The probability $p_{-1}$ that the observed value $Z(\omega)$ was
indeed caused by the nonprincipal solution is at most $\Prob\lbrace U
< -1/ | {\gamma}| \rbrace$, since $H_{\gamma}(u)$ changes its
monotonicity at $u = -1/\gamma$. 
For Gaussian $X$ and $\gamma= 0.1$---a very large value given
empirical evidence---this probability equals $7.62 \cdot10^{-24}$. For
an input with student $t$-distribution and $\nu= 4$ degrees of freedom
$p_{-1} \approx7.26 \cdot10^{-5}$. Hence, ignoring the nonprincipal
root to obtain unique latent data should not matter too much in practice.
\end{remark}

\begin{algorithm}[t]
\caption{Get input $\widehat{\mathbf
{x}}_{\tau}$: function \texttt{get.input($\cdot$)} in the \texttt
{LambertW} package.}\label{alg:get.input}

\begin{algorithmic}[1]
\REQUIRE data vector $\mathbf{y}$; parameter vector $\tau= (\mu_x,
\sigma_x, \gamma)$.
\ENSURE input vector $\widehat{\mathbf{x}}_{\tau}$.

\STATE$\mathbf{z} = (\mathbf{y} - \mu_x)/ \sigma_x$.
\STATE back-transform $\mathbf{z}$ via the principal branch to
$\mathbf{u} = W_{\gamma, 0}(\mathbf{z})$.
\RETURN$\widehat{\mathbf{x}}_{\tau} = \mathbf{u} \sigma_x +
\mu_x$.
\end{algorithmic}
\vspace*{4pt}
\end{algorithm}

Algorithm \ref{alg:get.input} describes the empirical version of
\eqref{eq:backtrafo_normalized_X}. The so obtained
%
%
\begin{equation}
\label{eq:back-transformed_x}
x_n = W_{\gamma, 0} \biggl(\frac{y_n - \mu_x}{\sigma_x} \biggr)
\sigma_x + \mu_x , \qquad n = 1, \ldots , N,
\end{equation}
is the input data $\widehat{\mathbf{x}}_{\tau}$ generating the
observed $\mathbf{y}$ and should have c.d.f. $F_X(x)$. Here $\widehat
{(\cdot)}$ does not stand for an estimate of $\tau$, but since $W_{\gamma,
0}(\cdot)$ ignores the nonprincipal branch, Algorithm \ref
{alg:get.input} need not return the ``true'' input data $\mathbf{x}$---even if $\tau$ is known.\footnote{This only applies if $p_{-1} =
\Prob ( U < -\frac{1}{| {\gamma}|}  ) > 0$, as otherwise
the back-transformation $W_{\gamma}(z) = W_{\gamma, 0}(z)$ is
bijective. In particular, if $U \geq0$---for example, for scale
family input $X \geq0$---then $\mathbf{x}_{\tau} \equiv\mathbf
{x}$, not just an approximation. See also Corollary \ref
{cor:cdf_pdf_scale_LambertW}.} For small $\gamma$, $\widehat{\mathbf
{x}}_{\tau}$ will most likely equal the true $\mathbf{x}$ for all
$n$; for large $\gamma$ some $y_j$'s might be falsely assigned to the
principal~$x_j$'s, although these $y_j$'s were actually caused by
nonprincipal $x_j$'s. For an estimate $\widehat{\tau}$ the notation
$\widehat{\mathbf{x}}_{\widehat{\tau}}$ will be used, which itself
is an approximation to~$\widehat{\mathbf{x}}_{\tau}$.

\section{Distribution and density function}
\label{sec:dist_dens}
For ease of notation and readability let
%
\begin{eqnarray}
\label{eq:define_z_u_x}
z &:=& \frac{y-\mu_x}{\sigma_x}, \qquad u_0 := W_{\gamma, 0}(z),
\qquad u_{-1} := W_{\gamma,-1}(z), \nonumber
\\[-8pt]
\\[-8pt]
x_{0} &:=& u_{0} \sigma_x + \mu_x, \qquad x_{-1} := u_{-1} \sigma_x +
\mu_x.
\nonumber
\end{eqnarray}

By definition,
\begin{eqnarray*}
G_Y(y) &=& \Prob(Y \leq y) = \Prob\bigl( \{ U \exp(\gamma U )
\} \sigma_x + \mu_x \leq y\bigr)\\
&=& \Prob ( U \exp{\gamma U} \leq z  ). 
\end{eqnarray*}

The transformation $H_{\gamma}(u)$ changes its monotonicity at $u =
-1/\gamma$ and its inverse $W_{\gamma, 0}(z)$ at $z = - 1/(\gamma
e)$. Consequently, the event $\lbrace Y < (>)\,\, y \rbrace$ [for $\gamma
>(<)\,\, 0$] has~to be split up into separate events in $U$ to derive the
distribution of $Y$.

\begin{theorem}[(Distribution of a location-scale $Y$)]
\label{theorem:cdf_Y}
The c.d.f. of a loca\-tion-scale Lambert $W\times F$ RV $Y$ equals (for
$\gamma> 0$)
%
%
\begin{equation}
\label{eq:cdf_LambertW_Y}
G_Y ( y \mid\bolds\beta, \gamma )=
\cases{\displaystyle
0,
&\quad  if   $\displaystyle y < - \frac{\sigma_x}{\gamma e} + \mu_x$,\vspace*{2pt}\cr\displaystyle
F_X ( x_0 \mid\bolds\beta ) - F_X (x_{-1} \mid\bolds\beta ),
&\quad  if   $\displaystyle-\frac{\sigma_x}{\gamma e} + \mu_x \leq y \leq\mu_x$, \vspace*{2pt}\cr\displaystyle
F_X ( x_0 \mid\bolds\beta ) ,
&\quad  if   $y \geq\mu_x$.}\hspace*{-28pt}
\end{equation}
The case $\gamma<0$ can be obtained analogously and for $\gamma= 0$
it is clear that $G_Y ( y \mid\bolds\beta, 0  ) =
F_X ( y \mid\bolds\beta )$.
\end{theorem}

\begin{pf}
Follows by matching the events in $Z$ with the corresponding events in
$U$ [\citet{Glen97}]; see Figure \ref{fig:Events_Z_Events_U}.
\end{pf}

For $z = -1/( \gamma e)$ both branches of $W(z)$ coincide, thus, $u_0 =
u_{-1}$. Therefore, $F_X(u_0 \sigma_x + \mu_x) - F_X(u_{-1} \sigma_x
+ \mu_x) \equiv0$ at $z = -1/(\gamma e)$, which implies continuity of
$G_Y(y)$ at $y = \mu_x -\frac{\sigma_x}{\gamma e}$; the same
reasoning shows continuity at $y = \mu_x$ ($z = 0$).

\begin{theorem}[(Density of a location-scale $Y$)]
\label{theorem:pdf_Y}
The p.d.f. of a location-scale Lambert $W\times F$ RV $Y$ equals (for
$\gamma> 0$)
%
\begin{equation}
\label{eq:pdf_LambertW_Y}
 \quad g_Y ( y \mid\bolds\beta, \gamma ) =
\cases{\displaystyle
0,  \qquad\hspace*{83.5pt}  \mbox{if   $\displaystyle y < - \frac{\sigma_x}{\gamma e} +
 \mu_x$,}\vspace*{4pt}\cr\displaystyle
f_X (x_0 \mid\bolds\beta ) \cdot W'_0 (\gamma z
 ) - f_X (x_{-1}\mid\bolds\beta ) \cdot W'_{-1}
(\gamma z  ), \vspace*{1pt}\cr
  \hspace*{115.5pt}\mbox{if }  \displaystyle- \frac{\sigma_x}{\gamma e} + \mu
_x \leq y \leq\mu_x, \vspace*{2pt}\cr\displaystyle
f_X (x_0 \mid\bolds\beta ) \cdot W'_0 (\gamma z
 ),  \qquad  \mbox{if   $y \geq\mu_x$.}
}
\end{equation}

Again, $\gamma<0$ can be obtained analogously, and $g_Y ( y \mid
\bolds\beta, 0  ) = f_X ( y \mid\bolds\beta
 )$.
\end{theorem}

\begin{pf}
Using that $\frac{d}{dz} W_{\gamma}(z) = W'(\gamma z)$, the first
derivative of $G_Y(y)$ with respect to $y$ equals \eqref{eq:pdf_LambertW_Y}.
The same arguments as for $G_Y(y)$ show that~$g_Y(y)$ is continuous at
$y = -\sigma_x/(\gamma e) + \mu_x$ and $y= \mu_x$.
\end{pf}

In general, the support of $Y$ depends on $\tau= \tau(\theta) \in T$
if $\gamma\neq0$. However, restricting $\tau$ to the subspace $S_{c}
:= \lbrace\tau\in T \mid-\sigma_x/(\gamma e) + \mu_x = c \rbrace$
gives the same support $[c, \infty)$ for all $\tau\in S_{c} \subseteq
T$ [or $(-\infty, c]$ for $\gamma<0$]. Of particular empirical
importance are
%
%
\begin{equation}
S_{0} := \lbrace\theta\in\Theta\mid\mu_x = \sigma_x/(\gamma e)
\rbrace \quad \mbox{and} \quad  S_{\pm\infty} = \lbrace\theta\in\Theta\mid
\gamma= 0 \rbrace.
\end{equation}

For (a scale family) $X \sim F_X ( x \mid\bolds\beta
)$ taking values in $[0, \infty)$ and $\gamma\geq0$, the support of
the corresponding (scale) Lambert $W\times F$ RV Y does not depend on
$\tau$ but always equals $[0, \infty)$.

\begin{corollary}[(C.d.f. and p.d.f. of a scale Lambert $W\times F$ RV)]
\label{cor:cdf_pdf_scale_LambertW}
If~$X$ is a nonnegative RV taking values in $[0, \infty)$ and $\gamma
\geq0$, then the inverse transformation $W_{\gamma}(y/\sigma_x)$ is
unique. Hence, the c.d.f. and p.d.f. of a scale Lambert $W\times F$ RV
$Y$ equal
%
%
\begin{equation}
\label{eq:cdf_LambertW_scale_Y}
G_Y ( y \mid\bolds\beta, \gamma ) = F_X \biggl(
W_{\gamma, 0} \biggl(\frac{y}{\sigma_x} \biggr) \sigma_x \Bigm|
\bolds\beta \biggr)
\end{equation}
and
%
%
\begin{equation}
\label{eq:pdf_LambertW_scale_Y}
g_Y ( y \mid\bolds\beta, \gamma ) = f_X
\biggl(W_{\gamma, 0} \biggl(\frac{y}{\sigma_x} \biggr) \cdot\sigma_x \Bigm|
\bolds\beta \biggr) \cdot W'_0 \biggl(\gamma\frac{y}{\sigma_x}
 \biggr).
\end{equation}
\end{corollary}

\begin{pf}
Follows by setting $\mu_x = 0$ in \eqref{eq:define_z_u_x}, \eqref
{eq:cdf_LambertW_Y} and \eqref{eq:pdf_LambertW_Y}, and noting that the
case $u < 0 $ does not exist since $X \geq0$.
\end{pf}

%
\begin{figure}
\includegraphics{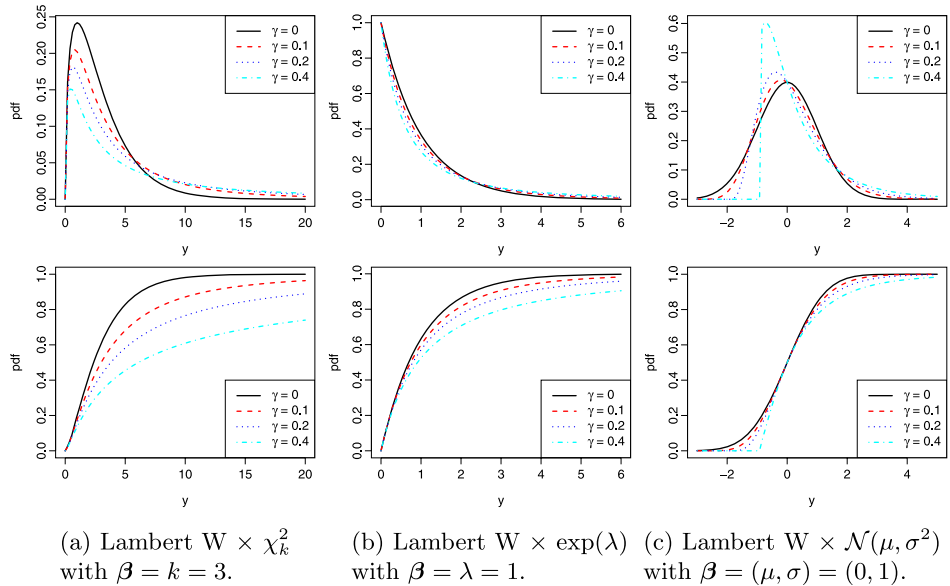}

\caption{The p.d.f. and c.d.f. of \textup{(a)} a ``noncentral, nonscaled,''
\textup{(b)} a
``scale'' and \textup{(c)}~a~``location-scale'' Lambert $W{}\times
{}$RV Y for
different degrees of skewness.}
\label{fig:LambertW_densities}
\end{figure}

For the c.d.f. and p.d.f. of a noncentral, nonscaled Lambert $W\times F$ RV
Y (Definition~\ref{def:noncentral_nonscaled_LambertW}) with $X$ taking
values in $[0, \infty)$ set $\sigma_x = 1$ in \eqref
{eq:cdf_LambertW_scale_Y} and \eqref{eq:pdf_LambertW_scale_Y}.

Theorems \ref{theorem:cdf_Y} and \ref{theorem:pdf_Y} demonstrate the
great flexibility of the Lambert~$W$ setting, since the closed form
expressions for $G_Y(y \mid\bolds\beta, \gamma)$ and $g_Y(y
\mid\bolds\beta, \gamma)$ hold for any well-defined input
$F_X(x \mid\bolds\beta)$ and $f_X(x \mid\bolds\beta)$,
respectively. Thus, researchers can easily create Lambert $W$ variants of
their favorite distribution $F_X$, by simply plugging $F_X$ and $f_X$
in \eqref{eq:cdf_LambertW_Y} and \eqref{eq:pdf_LambertW_Y}. Figure
\ref{fig:LambertW_densities} shows the p.d.f. and c.d.f. of the three
Lambert$\times F$ RVs discussed in Notation~\ref
{notation:LambertW_F} for four degrees of skewness, $\gamma= (0, 0.1,
0.2, 0.4)$. For $\gamma= 0$ the output~$Y$ equals the input $X$, thus,
also their p.d.f.s/c.d.f.s coincide (solid black lines). With increasing
$\gamma$, the RV $Y$---and thus its distribution and density---become
more and more skewed to the right (since $\gamma> 0$).

Although Lambert $W$ RVs are defined by transformation \eqref
{eq:location_scale_LambertW_Y}, they can be also considered as a
particular variant of an arbitrary $F_X$---independent of this
transformation. Sometimes the input/output aspect might be more
insightful (e.g., stock returns), whereas otherwise solely the
generalized distribution suffices to analyze a given data set.
Especially, if the latent variable~$X$ does not have any suitable
interpretation (see BMI data in Section~\ref{sec:applications}), one
can concentrate on the probabilistic properties of $Y$, ignoring the
input~$X$.

\subsection{Quantile function}
\label{subsec:quantile_general}
Equation \eqref{eq:cdf_LambertW_Y} and an inspection of Figure \ref
{fig:Events_Z_Events_U} directly relate $\mu_x$ to $Y$.
\begin{corollary}[(Median of $Y$)]
\label{cor:Y_median}
For a location-scale Lambert $W$ RV $Y$,
\[
\Prob(Y \leq\mu_x) = \Prob(X \leq\mu_x)  \qquad \mbox{for all } \gamma
\in\R.
\]
In particular, $\mu_x$ equals the median of $Y$, if $X$ is symmetric.
\end{corollary}

\begin{pf}
The transformation $H_{\gamma}(u) = u \exp(\gamma u)$ passes through
$(0,0)$ for all $\gamma\in\R$. Furthermore, $\exp(\gamma u) > 0$
for all $\gamma$ and all $u \in\R$. Therefore,
\begin{eqnarray*}
\Prob(Y \leq\mu_x) &=& \Prob(Z \leq0) = \Prob\bigl(U \exp(\gamma U)
\leq0\bigr) = \Prob(U \leq0) \\
&=& \Prob(X \leq\mu_x).
\end{eqnarray*}
For symmetric input $\Prob(X \leq\mu_x) = \frac{1}{2}$, therefore,
$\mu_x$ is the median of $Y$.
\end{pf}

Corollary \ref{cor:Y_median} not only gives a meaningful
interpretation of the parameter~$\mu_x$ for symmetric input, but the
sample median of $\mathbf{y}$ also yields a robust estimate of $\mu
_x$.

In general, the $\alpha$-quantile $y_{\alpha}$ of $Y$ satisfies
\begin{eqnarray*}
\alpha&\stackrel{!}{=}& \Prob(Y \leq y_{\alpha}) = \Prob\biggl(U \exp
(\gamma U) \leq\frac{y_{\alpha} - \mu_x}{\sigma_x}\biggr)\\
&=& \Prob\bigl(U \exp(\gamma U) \leq z_{\alpha}\bigr).
\end{eqnarray*}

For $\gamma> 0$ ($\gamma< 0$ analogously) and $z_{\alpha} > 0$ the
function $W_{\gamma,0}(\cdot)$ is bijective. Thus,
\begin{eqnarray*}
\Prob\bigl(U \exp(\gamma U) \leq z_{\alpha}\bigr) &=& \Prob\bigl(U \leq W_{\gamma
, 0}(z_{\alpha})\bigr) = \Prob\bigl(X \leq W_{\gamma, 0}(z_{\alpha}) \sigma
_x + \mu_x\bigr)
\end{eqnarray*}
and by definition of the $\alpha$-quantile of $X$,
%
%
\begin{equation}
\label{eq:x_alpha}
x_{\alpha} = W_{\gamma, 0}(z_{\alpha}) \sigma_x + \mu_x
 \quad \Leftrightarrow \quad  z_{\alpha} = u_{\alpha} e^{\gamma u_{\alpha}},
\end{equation}
where $u_{\alpha} = \frac{x_{\alpha} - \mu_x}{\sigma_x}$.

For $z_{\alpha} < 0$ and $\gamma> 0$, however, $W_{\gamma}(\cdot)$
is not bijective, thus, $z_{\alpha}$ cannot be computed explicitly as
in \eqref{eq:x_alpha}, but must be obtained by solving the implicit equation
\[
\alpha\stackrel{!}{=} F_X\bigl(W_{\gamma, 0}(z_{\alpha}) \sigma_x + \mu
_x\bigr) - F_X\bigl(W_{\gamma, -1}(z_{\alpha}) \sigma_x + \mu_x\bigr).
\]
In either case, the $\alpha$-quantile of $Y$ equals $y_{\alpha} =
z_{\alpha} \sigma_x + \mu_x$.

\section{Gaussian input}
\label{sec:Gaussian}
The results so far hold for any continuous input RV. To get a better
insight consider Gaussian input $U \sim\mathcal{N}(\mu_u, \sigma
_u^2)$ as a special case; here $\bolds\beta= (\mu_u, \sigma
_u)$. Its moment generating function $M_U(t)$ equals
\[
\E ( e^{t U}  ) = e^{t \mu_u +  {t^2}/{2} \sigma_u^2}
 \qquad \mbox{for all } t \in\R.
\]\vadjust{\eject}

Therefore, noncentral moments of $Z$ can be computed explicitly [see
\eqref{eq:non_central_mom_Z_derivative}] by
\[
\psi_{(n)} = n^{-n} \,\frac{\partial^n}{\partial\gamma^n} \exp
\biggl( \gamma n \mu_u + \gamma^2 \frac{n^2 \sigma_u^2}{2}  \biggr).
\]

In particular,
\begin{eqnarray*}
\mu_z &=& (\mu_u + \gamma\sigma_u^2) e^{\gamma\mu_u +
({\gamma^2}/{2})\sigma_u^2},\\[2pt]
\sigma_z^2 &=& 2^{-2} \bigl (e^{2 \gamma\mu_u + 2 \gamma^2 \sigma
_u^2} \bigl((4 \gamma\sigma_u^2 + 2 \mu_u)^2 + 4 \sigma_u^2\bigr)  \bigr) -
(\mu_u + \gamma\sigma_u^2)^2 e^{2 \gamma\mu_u + \gamma^2 \sigma
_u^2}\\[2pt]
&=& e^{2 \gamma\mu_u + 2 \gamma^2 \sigma_u^2} \bigl((2 \gamma\sigma
_u^2 + \mu_u)^2 + \sigma_u^2\bigr) - (\mu_u + \gamma\sigma_u^2)^2 e^{2
\gamma\mu_u + \gamma^2 \sigma_u^2}.
\end{eqnarray*}
As already mentioned in Section \ref{sec:LambertW_RV}, this is an
unstable system, in the sense that a small perturbation in $(\mu_u,
\sigma_u)$ results in a completely different $(\mu_z, \sigma_z)$ for
$\gamma\neq0$.

In contrast, the central moments of a location-scale Lambert $W{}\times
{}$Gaussian RV $Y$ with input $X \sim\mathcal{N}(\mu_x, \sigma_x^2)$
have a much simpler and stable form
%
%
\begin{equation}
\label{eq:mu_y_LambertW_Gaussian}
\mu_y = \mu_x + \sigma_x \E ( U e^{\gamma U}  )= \mu_x
+ \sigma_x \gamma e^{ {\gamma^2}/{2}},
\end{equation}
since $U = (X - \mu_x)/\sigma_x \sim\mathcal{N}(0, 1)$. Using
\eqref{eq:mu_y_LambertW_Gaussian}, the $k$th central moment of $Y$ can
be expressed by the $k$th central moment of $U e^{\gamma U}$,
\begin{eqnarray*}
\E(Y - \mu_y)^k 
= \sigma_x^k \E ( U e^{\gamma U} - \gamma e^{ {\gamma
^2}/{2}} )^k.
\end{eqnarray*}

In particular,
%
%
\begin{equation}
\label{eq:sigma_y_gamma_sigma_x}
\sigma_y^2
= \sigma_x^2 e^{\gamma^2}  \bigl((4 \gamma^2 +1) e^{\gamma^2} -
\gamma^2  \bigr),
\end{equation}
which only depends on the input variance and the skewness parameter
$\gamma$.

The main motive to introduce Lambert $W$ RVs is to accurately model
skewed data. The skewness coefficient of $Y$ is defined as $\gamma
_1(Y):=(\E(Y-\mu_y)^3)/\sigma_y^{3}$.\vspace*{-1pt} Analogously, the kurtosis
equals $\gamma_2(Y):=(\E(Y-\mu_y)^4)/\sigma_y^{4}$ and measures the
thickness of tails of $Y$. 

\begin{lemma}
\label{lem:skew_kurt_LambertW_Gaussian}
For a location-scale Lambert $W{}\times{}$Gaussian RV with input $X
\sim
\mathcal{N}(\mu_x, \sigma_x^2)$,
\begin{eqnarray}
\label{eq:skewness_LambertW_Gaussian}
\gamma_1(\gamma) = \gamma \biggl[ \frac{e^{3 \gamma^2}(9+27 \gamma
^2)- e^{\gamma^2}(3+12 \gamma^2)+5 \gamma^2}
{(e^{\gamma_2}(1+4 \gamma^2)-\gamma^2)^{3/2}}  \biggr]
\end{eqnarray}
and
\begin{eqnarray}
\label{eq:kurtosis_LambertW_Gaussian}
 \qquad \quad  \gamma_2(\gamma) = \frac{e^{6 \gamma^2} (3 +96 \gamma^2 + 256
\gamma^4) - e^{3 \gamma^2} (30 \gamma^2 + 60 \gamma^4 - 96 \gamma
^6)-3 \gamma^4}{ ( e^{\gamma^2}(1+4 \gamma^2)-\gamma^2  )^2}.
\end{eqnarray}
\end{lemma}

\begin{pf}
Dividing the third and fourth derivative of the moment generating
functions for a standard Gaussian $U$ at $t = \gamma n$ with respect to
$\gamma$ by~$n^{n}$ gives
\begin{eqnarray*}
\frac{1}{3^3} \frac{d^3}{d \gamma^3} e^{ {9 \gamma^2}/{2}}&=& 9
\gamma e^{ {9 \gamma^2}/{2}}+ 27 \gamma^3 e^{ {9 \gamma
^2}/{2}} = 9 \gamma e^{ {9 \gamma^2}/{2}}  ( 1 + 3 \gamma
^2 ), \\
\frac{1}{4^4} \frac{d^4}{d \gamma^4} e^{8 \gamma^2} &=& 3 e^{8
\gamma^2}+96 \gamma^2 e^{8 \gamma^2}+256 \gamma^4 e^{8 \gamma^2} =
e^{8 \gamma^2}  (3 +96 \gamma^2 +256 \gamma^4  ).
\end{eqnarray*}
The rest follows by expanding $\E (U e^{\gamma U} - \E U
e^{\gamma U}  )^3$ and $\E (U e^{\gamma U} - \E U e^{\gamma
U}  )^4$ via the binomial formula and using the above expressions.
\end{pf}

\begin{figure}

\includegraphics{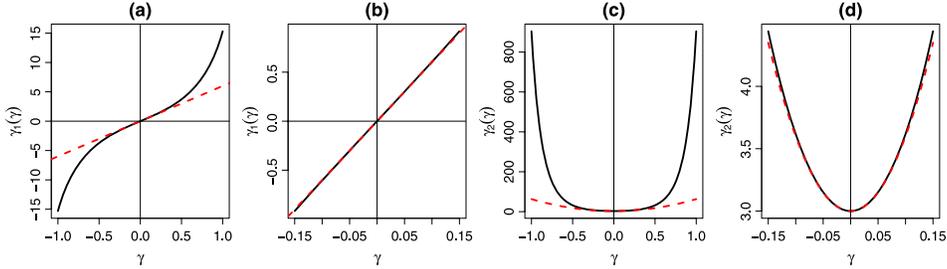}

\caption{Pearson skewness \textup{(a)} and kurtosis
\textup{(c)} coefficient for $\gamma\in[-1,1]$ and its first order Taylor
approximation (dashed line); \textup{(b)} and \textup{(d)}: zoom to
the interval $[-0.15,0.15]$.}\label{fig:skew_kurt_Y}
\vspace*{6pt}
\end{figure}

As expected, the skewness coefficient is an odd function in $\gamma$
with the same sign as $\gamma$. On the contrary, $\gamma_2(\gamma)$
is even. A first and second order Taylor approximation around $\gamma=
0$ yields $\gamma_1(\gamma) = 6 \gamma+ \mathcal{O}(\gamma^3)$ and
$\gamma_2(\gamma) = 3 + 60 \gamma^2+\mathcal{O}(\gamma^4)$,
respectively. Although $\gamma$ can take any value in $\R$, in
practice, it rarely exceeds $0.15$ in absolute value. In this interval
the Taylor approximation is almost indistinguishable from the true
function (Figure~\ref{fig:skew_kurt_Y}).\looseness=1

%

This first order approximation to $\gamma_1(\gamma)$ offers a rule of thumb
%
%
\begin{equation}
\label{eq:gamma_Taylor_rule}
\widehat{\gamma}_{\,\mathrm{Taylor}} := \frac{\widehat{\gamma}_1(\mathbf{y})}{6},
\end{equation}
which can be used as a starting value for better algorithms such as
 IGMM   and {MLE} (see Section \ref{sec:estimation}).

\begin{corollary}
\label{cor:limit_skew_kurt_LambertW_Gaussian}
The skewness and kurtosis coefficient are unbounded for $\gamma
\rightarrow\pm\infty$, that is,
\begin{eqnarray*}
\lim_{\gamma\rightarrow\pm\infty} \gamma_1(\gamma) = \pm\infty
 \quad \mbox{and} \quad  \lim_{\gamma\rightarrow\pm\infty} \gamma_2(\gamma)
= \infty.
\end{eqnarray*}
\end{corollary}

\begin{pf}
Omitting $-\gamma^2$ in the denominator and $5 \gamma^2$ in the
numerator of the skewness coefficient can be bounded from below
\begin{eqnarray*}
\frac{9 e^{3 \gamma^2}+27 \gamma^2 e^{3 \gamma^2}-3 e^{\gamma
^2}-12 \gamma^2 e^{\gamma^2}+5 \gamma^2}{(e^{\gamma^2}+4 \gamma^2
e^{\gamma^2}-\gamma^2)^{3/2}}
&\geq& \frac{e^{3 \gamma^2}(9 +27 \gamma^2)- e^{\gamma^2}(3+12
\gamma^2)}{e^{(3/2) \gamma^2}(1+4 \gamma^2)^{3/2}}\\
& = & e^{(3/2) \gamma^2}\frac{9 +27 \gamma^2}{(1+4 \gamma
^2)^{3/2}}\\
& &{}- e^{-\gamma^2 / 2} \frac{3+12 \gamma^2}{(1+4 \gamma^2)^{3/2}}.
\end{eqnarray*}
As the exponential function dominates rational functions, the first
term tends to~$\infty$, whereas the second one goes to $0$ for $\gamma
$ to $\infty$.

In case of the kurtosis coefficient, the term $e^{6 \gamma^2}$ in the
numerator dominates all other terms for large $\gamma$ and thus
determines the asymptotic behavior of $\gamma_2(\gamma)$ for $\gamma
$ to $\pm\infty$.
\end{pf}

This result shows that the Lambert $W{}\times{}$Gaussian distributions
can be used to model a larger variety of skewed data than a skew-normal
distribution, since its skewness coefficient is restricted to the
interval $(-0.995, 0.995)$ [\citet{Azzalini85}].

\section{Parameter estimation}
\label{sec:estimation}
For a sample of $N$ independent identically distributed (i.i.d.)
observations $\mathbf{y}=( y_1, \ldots , y_N )$, which presumably
originates from transformation \eqref{eq:location_scale_LambertW_Y},
$\theta= (\bolds\beta, \gamma)$ has to be estimated from the data.
In addition to the commonly used maximum likelihood estimator ({MLE}) for~$\theta$,
I also present a method of moments estimator for
$\tau$ that builds on the input/output relation in Figure \ref
{fig:LambertW_flowchart}.

\subsection{Maximum likelihood estimation}
\label{sec:LambertW_MLE}
The log-likelihood function in the i.i.d. case equals
%
%
\begin{equation}
\label{eq:likelihood_general}
\ell ( \theta\mid\mathbf{y}  ) = \sum_{i=1}^{N} \log
g_Y (y_i \mid\theta ),
\end{equation}
where $g_Y (\cdot\mid\theta )$ is the p.d.f. of $Y$---see
\eqref{eq:pdf_LambertW_Y}. The {MLE} is that $\theta=
(\bolds\beta, \gamma)$ which maximizes the log-likelihood
\[
\widehat{\theta}_{\mathrm{MLE}} = \arg\max_{\theta} \ell (
\theta\mid\mathbf{y}  ) .
\]

Since $g_Y(y_i \mid\theta)$ is a function of $f_X(x_i \mid
\bolds\beta)$, the  {MLE} depends on the specification of
the input density. In general, this multivariate, nonlinear
optimization problem must be carried out by numerical methods, as the
two branches of $W(z)$ for $y \leq\mu_x$ do not allow any further
simplification.

For (scale) Lambert $W \times F_X$ with support in $(0, \infty)$
and $\gamma\geq0$, however, $g_Y ( y \mid\bolds\beta,
\gamma ) = f_X (W_{\gamma, 0} (\frac{y}{\sigma
_x} ) \cdot\sigma_x \mid\bolds\beta ) \cdot
W'_0 (\gamma\frac{y}{\sigma_x}  )$ (Corollary \ref
{cor:cdf_pdf_scale_LambertW}). Thus, \eqref{eq:likelihood_general} can
be rewritten as
%
%
\begin{eqnarray}
\label{eq:loglikelihood_Y_with_xi}
\ell (\bolds\beta, \gamma\mid\mathbf{y}  ) &=&
\ell \bigl( \bolds\beta\mid\mathbf{x}_{(0, \sigma_x, \gamma
)} \bigr) + \sum_{i=1}^{N} \log W'_0 \biggl(\gamma\frac{y_i}{\sigma
_x}  \biggr),
\end{eqnarray}
where
%
%
\begin{equation}
\label{eq:likelihood_x.hat}
\ell\bigl ( \bolds\beta\mid\mathbf{x}_{(0, \sigma_x, \gamma
)} \bigr) = \sum_{i=1}^{N} \log f_X \biggl(W_{\gamma, 0} \biggl(\frac
{y_i}{\sigma_x} \biggr) \cdot\sigma_x \Bigm|\bolds\beta \biggr)
\end{equation}
is the log-likelihood of the back-transformed data $\mathbf{x}_{(0,
\sigma_x, \gamma)} = W_{\gamma, 0} (\frac{\mathbf{y}}{\sigma
_x} ) \cdot\sigma_x$ [no $\widehat{(\cdot)}$ since the
inverse is unique in this case].
Note that $W'_0 (\gamma\frac{y_i}{\sigma_x}  )$ only
depends on $\sigma_x(\bolds\beta)$ (and $\gamma$), but not
necessarily on every coordinate of $\bolds\beta$.

The equivalence \eqref{eq:loglikelihood_Y_with_xi} shows the relation
between the exact MLE $  (\widehat{\bolds\beta}, \widehat
{\gamma} )$ based on $\mathbf{y}$ and the approximate MLE
$\widehat{\bolds\beta}$ based on $\mathbf{x}_{(0, \sigma_x,
\gamma)}$: if we would know $\sigma_x$ and $\gamma$ beforehand, then
we could just back-transform $\mathbf{y}$ to $\mathbf{x}_{(0, \sigma
_x, \gamma)}$ and compute $\widehat{\bolds\beta}_{\mathrm{MLE}}$ based
on $\mathbf{x}_{(0, \sigma_x, \gamma)}$ [maximize \eqref
{eq:likelihood_x.hat}]; however, in practice, $\sigma_x$ and $\gamma$
have to be estimated from $\mathbf{y}$ and this uncertainty enters the
log-likelihood \eqref{eq:loglikelihood_Y_with_xi} by the additional
term $\sum_{i=1}^{n} \log W'_0 (\gamma\frac{y_i}{\sigma_x}
 )$.

For $z > 0$ it can be easily shown that $W'(z) = \frac{W(z)}{z (1 +
W(z))} > 0$ as well as $W'(z) < 1$ since $W'(0) = 1$ and $W''(z) =
-W'(z) \exp(-W(z)) \frac{W(z) + 2}{(W(z)+1)^2} < 0$. Hence, $\sum
_{i=1}^{n} \log W'_0 (\gamma\frac{y_i}{\sigma_x}  ) < 0$
for $\gamma> 0$ and can be thought of as a penalty for transforming
$\mathbf{y}$ to the ``nicer'' $\mathbf{x}_{(0, \widehat{\sigma}_x,
\widehat{\gamma})}$ with estimated parameters: the larger $\gamma$,
the bigger the penalty on the log-likelihood $\ell ( \bolds
\beta\mid\mathbf{x}_{(0, \sigma_x, \gamma)} )$ of the
``nice'' back-transformed data, since $W''(z) < 0$. 

\subsubsection*{Parameter-dependent support}
For location-scale Lambert $W\times F$ RVs the support of $g_Y(y)$
depends on $\tau= \tau(\theta)$ and therefore violates a crucial
assumption of most results related to (asymptotic) properties of the
MLE. 
Only for $\gamma= 0$ the support of $g_Y(y) = f_X(y)$ does not depend
on $\theta$. For $X \sim\mathcal{N}(0,1)$ it can be shown that the
Fisher information matrix $I_N(\gamma) = - \E ( \frac{d^2}{d
\gamma^2} \ell (\theta\mid\mathbf{y}  )  ) = 8
N$. Hence, for the symmetric Gaussian case $\sqrt{N} \widehat{\gamma
}_{\mathrm{MLE}} \rightarrow\mathcal{N} (0, \frac{1}{8} )$.
Simulations in Section \ref{sec:simulations} confirm this asymptotic
result and suggest that also for the general Gaussian case $\widehat
{\theta}_{\mathrm{MLE}}$ is well behaved, that is, it is $\sqrt
{N}$-consistent and asymptotically efficient.

A theoretical analysis of the asymptotic behavior of the MLE for
$\gamma\neq0$ is beyond the scope of this study, but simulations show
that also for parameter dependent support $\widehat{\theta}_{\mathrm
{MLE}}$ is an unbiased estimator with root mean square errors
comparable to the $\gamma = 0$ case.

\subsection{Iterative generalized method of moments (IGMM)}
\label{sec:IGMM}
A disadvantage of the  {MLE} is the mandatory a-priori
specification of the input distribution. In practice, however, it is
rarely known what kind of distribution is a good fit to the data, even
more so if the data is transformed via a nonlinear transformation.
Thus, here I present an iterative method to estimate the optimal
inverse-transformation \eqref{eq:backtrafo_normalized_X} by estimating
$\tau$ directly, instead of estimating~$\theta$ and then computing
$\tau(\widehat{\theta})$. This method builds on the input/output
aspect and only relies upon the specification of the theoretical
skewness of $X$.

The proposed estimator for $\tau$ works as follows (see below for a
more detailed discussion):
\begin{longlist}[(3)]
\item[(1)] set starting values $\tau^{(0)} = \tau_0$. Set $k=0$;
\item[(2)]\label{item:gamma} assume $\mu_x^{(k)}$ and $\sigma_x^{(k)}$
are known and estimate $\gamma$ from\vspace*{-2pt} $\mathbf{z}^{(k)} = \frac
{\mathbf{y} - \mu_x^{(k)}}{\sigma_x^{(k)}}$ to obtain $\gamma
^{(k+1)}$ (Algorithm \ref{alg:gamma_GMM});
\item[(3)]\label{item:mu_sigma} assume $\gamma^{(k+1)}$ is known and get
estimates $\mu_x^{(k+1)}$ and $\sigma_x^{(k+1)}$ from the
back-transformed data $\widehat{\mathbf{x}}_{(\mu_x^{(k)}, \sigma
_x^{(k)}, \gamma^{(k+1)})}$ (Algorithm \ref{alg:IGMM}). Set $k = k+1$;
\item[(4)] iterate between (\ref{item:gamma}) and (\ref{item:mu_sigma}) until
convergence of the sequence $\tau^{(k)}$.
\end{longlist}

\begin{algorithm}[b]
\caption{Find optimal $\gamma$: function
\texttt{gamma\_GMM($\cdot$)} in the \texttt{LambertW} package.}\label{alg:gamma_GMM}

\begin{algorithmic}[1]
\REQUIRE standardized data vector $\mathbf{z}$; theoretical skewness
$\gamma_1(X)$.
\ENSURE$\widehat{\gamma}_{\mathrm{GMM}}$ as in \eqref{eq:gamma_GMM}.

\STATE Compute lower and upper bound for $\gamma$: $\mathit{lb} = -\frac
{1}{\exp(1) \max(\mathbf{z})}$\vspace*{-1pt} and $\mathit{ub} = -\frac{1}{\exp(1) \min
(\mathbf{z})}$.
\STATE$\widehat{\gamma}_{\mathrm{GMM}} = \arg\min_{\gamma} \|\widehat
{\gamma}_1(\widehat{\mathbf{u}}) - \gamma_1(X)\|$ where
$\widehat{\mathbf{u}} = W_{\gamma}(\mathbf{z})$ subject to $\gamma
\in[\mathit{lb},\mathit{ub}]$.
\RETURN$\widehat{\gamma}_{\mathrm{GMM}}$.
\end{algorithmic}
\end{algorithm}

For a moment assume that $\mu_x$ and $\sigma_x$ are known and only
$\gamma$ has to be estimated. Since $\mu_x$ and $\sigma_x$ are
known, we can consider $\mathbf{z} = \frac{\mathbf{y} - \mu
_x}{\sigma_x}$. A natural\vspace*{1pt} choice for $\gamma$ is the one that results
in back-transformed data $\widehat{\mathbf{u}}_{\gamma} = W_{\gamma
} ( \mathbf{z}  )$ with sample skewness equal to the
theoretical skewness of $U$, which equals the theoretical skewness of~$X$. Formally,
%
%
\begin{equation}
\label{eq:gamma_GMM}
\widehat{\gamma}_{\mathrm{GMM}} = \arg\min_{\gamma} \|\gamma
_1(X) - \widehat{\gamma}_1(\widehat{\mathbf{u}}_{\gamma})\|,
\end{equation}
where $\|\cdot\|$ is a proper norm in $\R$, for example,
$\|s\| = s^2$ or $\|s\| =  | s  |$.

\subsubsection*{\texorpdfstring{Discussion of Algorithm \protect\ref{alg:gamma_GMM}}{Discussion of Algorithm 2}}

For example, let $\mathbf{y}$ be positively skewed data, $\widehat
{\gamma}_1(\mathbf{y}) > 0$, and the input $\mathbf{x}$ causing the
observed $\mathbf{y}$ is assumed/known to be symmetric, thus, $\gamma
_1(X)=0$. By the nature of transformation $H_{\gamma}(u)$, the
skewness parameter $\gamma$ must be also positive and the Taylor
approximation of $\gamma_1(\gamma)$ for Gaussian input [see \eqref
{eq:gamma_Taylor_rule}] gives a good initial estimate $\gamma_0 =
\widehat{\gamma}_1(\mathbf{y})/6 > 0$. In the same way as the
mapping $u \mapsto u \exp(\gamma u)$ introduces skewness, the inverse
transformation $W_{\gamma}(\mathbf{z})$ results in less positively
skewed~$\widehat{\mathbf{u}}_{\gamma}$ due to the curvature in
$W_{\gamma}(\cdot)$ (see Figure \ref{fig:Events_Z_Events_U}). As the
initial guess~$\gamma_0$ rarely gives exactly symmetric input,
Algorithm \ref{alg:gamma_GMM} searches for a $\gamma$ such that the
empirical skewness of $\widehat{\mathbf{u}}_{\gamma}$ is as close as
possible to the ``true'' skewness~$\gamma_1(X)$.

There are natural bounds for $\gamma$ to guarantee the observability
of $\mathbf{y}$, for example, a $\gamma$ too large makes large
negative observations in $\mathbf{y}$ impossible (due to the minimum
at $z = -1/e$; see Figure \ref{fig:Events_Z_Events_U}). However,
since $\mathbf{y}$ has actually been observed, the search space for
$\gamma$ must be limited to the interval $O_{\mathbf{z}} := [-\frac
{1}{\exp(1) \max(\mathbf{z})}, -\frac{1}{\exp(1) \min(\mathbf
{z})}]$. If there exists a $\tilde{\gamma} \in O_{\mathbf{z}}$ such
that $\widehat{\gamma}_1(\widehat{\mathbf{u}}_{\tilde{\gamma}}) =
\gamma_1(X)$, then Algorithm \ref{alg:gamma_GMM} will return
$\widehat{\gamma} = \tilde{\gamma}$ due to the monotonically
increasing curvature of $H_{\gamma}(u)$ and $W_{\gamma}(z)$
respectively; if there is no such $\tilde{\gamma} \in O_{\mathbf
{z}}$, then Algorithm \ref{alg:gamma_GMM} returns either the lower or
upper bound of~$O_{\mathbf{z}}$, depending on whether $\mathbf{z}$ is
negatively or positively skewed.

This univariate minimization problem with constraints can be carried
out by standard optimization algorithms.

In practice, $\mu_x$ and $\sigma_x$ are rarely known but also have to
be estimated from the data. As $\mathbf{y}$ is shifted and scaled
\textit{ahead of} the back-transformation~$W_{\gamma, 0}(\cdot)$, the
initial choice of $\mu_x$ and $\sigma_x$ affects the optimal choice
of $\gamma$. Therefore, the optimal triple $(\widehat{\mu}_x,
\widehat{\sigma}_x, \widehat{\gamma})$ must be obtained
iteratively. 

\subsubsection*{\texorpdfstring{Discussion of Algorithm \protect\ref{alg:IGMM}}{Discussion of Algorithm 3}}
Algorithm \ref
{alg:IGMM} first computes $\mathbf{z}^{(k)} = (\mathbf{y} - \mu
_x^{(k)})/\allowbreak\sigma_x^{(k)}$ using $\mu_x^{(k)}$ and $\sigma_x^{(k)}$
from the previous step. This normalized output can then be passed to
Algorithm \ref{alg:gamma_GMM} to obtain an updated $\gamma^{(k+1)}:=
\widehat{\gamma}_{\mathrm{GMM}}$. Using this new $\gamma^{(k+1)}$, one can
back-transform $\mathbf{z}^{(k)}$ to the presumably zero-mean,
unit-variance input $\mathbf{u}^{(k+1)} = W_{\gamma^{(k+1)}}(\mathbf
{z}^{(k)})$. Herewith we can obtain a~better approximation to the
``true'' latent $\mathbf{x}$ by $\mathbf{x}^{(k+1)} = \mathbf
{u}^{(k+1)} \sigma_x^{(k)} + \mu_x^{(k)}$. However, $\gamma
^{(k+1)}$---and therefore $\mathbf{x}^{(k+1)}$---has been obtained
using $\mu_x^{(k)}$ and $\sigma_x^{(k)}$ which are not necessarily
the most accurate estimates in light of the updated approximation
$\widehat{\mathbf{x}}_{( \mu_x^{(k)}, \sigma_x^{(k)} , \gamma
^{(k+1)})}$. Thus, Algorithm \ref{alg:IGMM} computes new estimates
$\mu_x^{(k+1)}$ and $\sigma_x^{(k+1)}$\vspace*{-1pt} by the sample mean and
standard deviation of $\widehat{\mathbf{x}}_{( \mu_x^{(k)}, \sigma
_x^{(k)} , \gamma^{(k+1)})}$,\vspace*{-3pt} and starts another iteration by passing
the updated normalized output $\mathbf{z}^{(k+1)} = \frac{\mathbf{y}
- \mu_x^{(k+1)}}{\sigma_x^{(k+1)}}$ to Algorithm \ref{alg:gamma_GMM}
to obtain a new $\gamma^{(k+2)}$.

\begin{algorithm}[t]
\caption{Iterative generalized method of moments:
function \texttt{IGMM($\cdot$)} in the \texttt{LambertW} package.}\label{alg:IGMM}

\begin{algorithmic}[1]
\REQUIRE data vector $\mathbf{y}$; tolerance level $\mathit{tol}$; theoretical
skewness $\gamma_1(X)$.
\ENSURE IGMM parameter estimate $\widehat{\tau}_{\mathrm{IGMM}} =
(\widehat{\mu}_x, \widehat{\sigma}_x, \widehat{\gamma})$.

\STATE Set $\tau^{(-1)} = (0,0,0)$.
\STATE Starting values: $\tau^{(0)} = (\mu_x^{(0)}, \sigma
_x^{(0)},\gamma^{(0)})$, where $\mu_x^{(0)} = \tilde{\mathbf{y}}$
and $\sigma_x^{(0)} = \overline{\sigma}_y$ are the sample median and
standard deviation of $\mathbf{y}$, respectively. $\gamma^{(0)} =
\frac{\widehat{\gamma}_1(\mathbf{y}) - \gamma_1(X)}{6}$
$\rightarrow$ see \eqref{eq:gamma_Taylor_rule} for details.
\STATE$k=0$.
\WHILE{$\|\tau^{(k)} - \tau^{(k-1)}\| > \mathit{tol}$}
\STATE$\mathbf{z}^{(k)} = (\mathbf{y} - \mu_x^{(k)})/\sigma_x^{(k)}$,\vspace*{1pt}
\STATE Pass $\mathbf{z}^{(k)}$ to Algorithm \ref{alg:gamma_GMM}
$\longrightarrow\gamma^{(k+1)}$,\vspace*{1pt}
\STATE back-transform $\mathbf{z}^{(k)}$ to $\mathbf{u}^{(k+1)} =
W_{\gamma^{(k+1)}}(\mathbf{z}^{(k)})$; compute $\mathbf{x}^{(k+1)} =
\mathbf{u}^{(k+1)} \sigma_x^{(k)} + \mu_x^{(k)}$,
\STATE update parameters: $\mu_x^{(k+1)} = \overline{\mathbf
{x}}_{k+1}$ and $\sigma_x^{(k+1)} = \widehat{\sigma}_{x_{k+1}}$,
\label{line:set_new_theta}
\STATE$\tau^{(k+1)} = ( \mu_x^{(k+1)}, \sigma_x^{(k+1)},
\gamma^{(k+1)})$,
\STATE$k=k+1$.
\ENDWHILE
\RETURN$\tau_{\mathrm{IGMM}} = \tau^{(k)}$.
\end{algorithmic}
\end{algorithm}

The algorithm returns the optimal $\widehat{\tau}_{\mathrm{IGMM}}$
once the estimated parameter triple does not change anymore from one
iteration to the next, that is, if $\|\tau^{(k)} - \tau ^{(k+1)}\| < \mathit{tol}$.

A great advantage of the  {IGMM} estimator is that it does not
require any further specification of the input except its skewness. For
example, no matter if the input is normally, student-$t$, Laplace or
uniformly distributed, the  {IGMM} estimator finds a $\tau$
that gives symmetric $\widehat{\mathbf{x}}_{\widehat{\tau}}$
independent of the particular choice of (symmetric) $F_X(\cdot)$. 

A disadvantage of  {IGMM} from a probabilistic point of view is
its determination. In general, Algorithm \ref{alg:IGMM} will lead to back-transformed
data with sample skewness identical to $\gamma_1(X)$ and so no
stochastic element remains in the nature of the estimator.\footnote{If
$\gamma_1(X)$ depends on one or more parameters of the distribution of
$X$ (e.g., Gamma), then the  {IGMM} algorithm must be adapted
to this very problem.} Note that  {IGMM} does not provide an
estimate of $\bolds\beta$ (except for Gaussian input); if
necessary, an estimate of $\bolds\beta$ must be obtained in a
separate step, for example, by estimating $\bolds\beta$ from the
back-transformed data $\widehat{\mathbf{x}}_{\widehat{\tau}}$.
However,\vspace*{1pt} in general, $\widehat{\bolds\beta}_{\mathrm{MLE}}$ estimated
only from $\widehat{\mathbf{x}}_{\widehat{\tau}}$ is (slightly)
different from $\widehat{\bolds\beta}_{\mathrm{MLE}}$ using Lambert $W$
MLE on the original data~$\mathbf{y}$: in the first case $\widehat
{\tau}$ is assumed to be known and fixed, whereas in the second case
$\bolds\beta$ and $\tau$ are estimated jointly [see \eqref
{eq:loglikelihood_Y_with_xi}].

The underlying input data $\mathbf{x} = (x_1, \ldots , x_N)$ can be
approximated via Algorithm~\ref{alg:get.input} using $\widehat{\tau
}_{\mathrm{IGMM}}$. The so obtained $\widehat{\mathbf{x}}_{\widehat{\tau}_{\mathrm
{IGMM}}}$ may then be used to check if $X$ has characteristics of a
known parametric distribution $F_X(x \mid\bolds\beta)$, and
thus is an easy, but heuristic check if $\mathbf{y}$ follows a
particular Lambert $W \times F_X$ distribution. However, such a test
can only serve as a rule of thumb for various reasons: (i) $\widehat
{\tau} \neq\tau$, thus tests are too optimistic as $\widehat
{\mathbf{x}}_{\widehat{\tau}}$ will have ``nicer'' properties
regarding $F_X$ than the true $\mathbf{x}$ would have; (ii) ignoring
the nonprincipal branch alters the sample distribution of the input---putting
no observations to the far left (or right): not so much of a
problem for small~$\gamma$, the distribution can be truncated
considerably for large $\gamma$. For Gaussian input various tests are
available [Jarque--Bera, Shapiro--Wilk, among others; see \citet
{Thode02}], for other distributions a Kolmogorov--Smirnov test can be
used.\footnote{If the data does not represent an independent sample
(as usual for financial data), then critical values of several test
statistics need not be valid anymore and adapted tests should be used
[see \citet{Weiss78}].}

\subsubsection{Gaussian IGMM}
For Gaussian $X$ the system of equations
%
%
\begin{eqnarray}
\label{eq:mu_y_Gaussian}
\mu_y(\gamma) &=& \mu_x + \sigma_x \gamma e^{ {\gamma
^2}/{2}},\\
\label{eq:sigma2_y_Gaussian}
\sigma_y^2(\gamma) &=& \sigma_x^2 e^{\gamma^2}  \bigl((4 \gamma^2
+1) e^{\gamma^2} - \gamma^2  \bigr)
\end{eqnarray}
has a unique solution for $(\mu_x, \sigma_x)$. Given $\widehat
{\gamma}_{\mathrm{IGMM}}$ and the sample moments~$\overline{\mu}_y$
and $\overline{\sigma}_y$, the input parameters $\mu_x$ and $\sigma
_x^2$ can be obtained by
%
%
\begin{eqnarray}
\label{eq:IGMM_sigma2}
\widehat{\sigma}_x^2(\widehat{\gamma}_{\mathrm{IGMM}}) &=& \frac
{\overline{\sigma}_y^2}{e^{\widehat{\gamma}_{\mathrm{IGMM}}^2}
 ((4 \widehat{\gamma}_{\mathrm{IGMM}}^2 +1) e^{\widehat{\gamma
}_{\mathrm{IGMM}}^2} - \gamma_{\mathrm{IGMM}}^2  )},\\
\label{eq:IGMM_mu}
\widehat{\mu}_x(\widehat{\gamma}_{\mathrm{IGMM}}) &=& \overline
{\mu}_y - \widehat{\sigma}_x^2(\widehat{\gamma}_{\mathrm{IGMM}})
\widehat{\gamma}_{\mathrm{IGMM}} e^{ {\widehat{\gamma
}_{\mathrm{IGMM}}^2}/{2}}.
\end{eqnarray}
Hence, line \ref{line:set_new_theta} of Algorithm \ref{alg:IGMM} can
be altered to
 %
\begin{equation}
\begin{tabular}{@{}p{300pt}@{}}
\ref{line:set_new_theta}b: $\mu_x^{(k+1)} = \widehat{\mu
}_x(\overline{\mu}_y,\overline{\sigma}_y, \gamma_{k+1})$ and
$\sigma_x^{(k+1)}= \widehat{\sigma}_x(\overline{\sigma}_y, \gamma
_{k+1})$, given by \eqref{eq:IGMM_sigma2} and \eqref{eq:IGMM_mu}.
\end{tabular}
\end{equation}

\begin{algorithm}
\caption{Random sample generation: function
\texttt{rLambertW($\cdot$)} in the \texttt{LambertW} package.}\label{alg:sim_LambertW}
\begin{algorithmic}[1]
\REQUIRE number of observations $n$; parameter vector $\bolds
\beta$; specification of the input distribution $F_X(x \mid
\bolds\beta)$; skewness parameter $\gamma$.
\ENSURE random sample $(y_1, \ldots , y_n)$ of a Lambert $W \times
 F$ RV.

\STATE Simulate $n$ samples $\mathbf{x} = (x_1, \ldots , x_n) \sim
F_X(x \mid\bolds\beta)$.
\STATE Compute $\mu_x(\bolds\beta)$ and $\sigma_x(\bolds
\beta)$ given the type of Lambert $W\times F$ distribution
(noncentral, nonscale; scale; location-scale).
\STATE$\mathbf{u} = (\mathbf{x}-\mu_x(\bolds\beta)) / \sigma
_x(\bolds\beta)$.
\STATE$\mathbf{z} = \mathbf{u} \exp(\gamma\mathbf{u})$.
\RETURN$\mathbf{y} = \mathbf{z} \sigma_x(\bolds\beta) +
\mu_x(\bolds\beta)$.
\end{algorithmic}
\end{algorithm}

Even though this simplification would lead to a faster estimation of
$\tau$, it is mostly of theoretical interest, as it cannot be
guaranteed that $X$ indeed is Gaussian; 
the more general Algorithm \ref{alg:IGMM} should be used in
practice.\footnote{All numerical estimates $\widehat{\tau}_{\mathrm
{IGMM}}$ reported in Section \ref{sec:simulations} were obtained using
the more general algorithm with line \ref{line:set_new_theta}, not
\ref{line:set_new_theta}b.}

\section{Simulations}
\label{sec:simulations}
Although the c.d.f., p.d.f. and moments of a Lambert~$W$ RVs are nontrivial
expressions, their simulation is straightforward (Algorithm \ref
{alg:sim_LambertW}).

This section explores the finite-sample properties of estimators for
$\theta= (\mu_x, \sigma_x, \gamma)$ under Gaussian input $X \sim
\mathcal{N}(\mu_x, \sigma_x^2)$.\footnote{For the special case of
Gaussian input $\tau\equiv\theta$, thus,  {IGMM} estimates
$\widehat{\tau}_{\mathrm{IGMM}} = \widehat{\theta}_{\mathrm{IGMM}}$ can be
compared directly to $\widehat{\theta}_{\mathrm{MLE}}$.} In
particular, conventional Gaussian  {MLE} (estimation of $\mu_y$
and $\sigma_y$ only; $\gamma\equiv0$),  {IGMM} and Lambert
$W{}\times{}$Gaussian  {MLE}, and---for a skew competitor---the
skew-normal MLE\footnote{Function \texttt{sn.mle} in the R package
\texttt{sn}.} are studied. Whereas a comparison of accuracy and
efficiency in $\widehat{\gamma}$ does not make sense, it is
meaningful to analyze $\widehat{\mu}_y$ and $\widehat{\sigma}_y$ of
skew-normal versus Lambert $W{}\times{}$Gaussian MLE.

\subsection*{Scenarios} Each estimator is applied to 3 kinds of
simulated data sets for $4$ different sample sizes of $N=50, 100, 250$
and $1\mbox{,}000$:
\begin{longlist}
\item[$\gamma= 0$:] Data is sampled from a symmetric RV $Y = X \sim
\mathcal{N}(0,1)$. Does additional estimation of $\gamma$ affect the
properties of $\widehat{\mu}_y$ or $\widehat{\sigma}_y$?
\item[$\gamma= -0.05$:] A typical value for financial data, such as
the LATAM returns introduced in Section \ref{sec:introduction}.
\item[$\gamma= 0.3$:] This large value reveals the importance of the
two branches of the Lambert $W$ function. How does the skew-normal MLE
handle extremely skewed data [$\gamma(0.3) = 1.9397$]?
\end{longlist}

Simulations are based on $n=1\mbox{,}000$ replications. The input mean $\mu_x$
and standard deviation $\sigma_x$ are chosen such that the observed RV
has $\mu_y(\gamma) = 0$ and $\sigma_y(\gamma)=1$ for all $\gamma$.
These functional relations can be obtained by \eqref{eq:IGMM_sigma2}
and \eqref{eq:IGMM_mu}. For IGMM the tolerance level was set to $\mathit{tol} =
10^{-6}$ and the Euclidean norm was used.

\begin{remark}
The Gaussian and skew-normal MLE estimate the mean and standard
deviation of $Y$. Both Lambert $W$ methods estimate the mean and standard
deviation of the latent variable $X$ plus the skewness parameter~$\gamma$.
Thus, for a meaningful comparison the implied estimates
$\widehat{\sigma}_y(\widehat{\mu}_x, \widehat{\gamma})$ and
$\widehat{\mu}_y(\widehat{\mu}_x, \widehat{\sigma}_x, \widehat
{\gamma})$ given by \eqref{eq:mu_y_Gaussian} and \eqref
{eq:sigma2_y_Gaussian} are reported below.
\end{remark}

\subsection{\texorpdfstring{Symmetric data: $\gamma= 0$}{Symmetric data: gamma = 0}}
This parameter choice investigates if imposing the Lambert $W$ framework,
even though its use is superfluous, causes a~quality loss in the
estimation. Furthermore, critical values can be obtained for the finite
sample behavior of $\widehat{\gamma}$ under the null hypothesis of a
symmetric distribution.\vadjust{\eject}

%
\begin{table}
\tabcolsep=0pt
\caption{Bias and  {RMSE} of $\widehat
{\theta}$ for $\gamma= 0$ and $X \sim N(0,1)$}\label{tab:N01_sim}
\begin{tabular*}{\textwidth}{@{\extracolsep{\fill}}ld{5.0}d{2.4}d{2.4}d{2.4}d{1.4}d{1.4}d{1.4}@{}}
\hline
 &&\multicolumn{3}{c}{\textbf{Bias}}&\multicolumn{3}{c@{}}{\textbf{RMSE}${}\bolds{\cdot}{}\bolds{\sqrt{N}}$}\\[-5pt]
&&\multicolumn{3}{c}{\hrulefill}&\multicolumn{3}{c@{}}{\hrulefill}\\
 &\multicolumn{1}{c}{$\bolds{N}$}&\multicolumn{1}{c}{$\bolds{\gamma=0}$}&\multicolumn{1}{c}{$\bolds{\mu_y = 0}$}
 &\multicolumn{1}{c}{$\bolds{\sigma_y = 1}$} &\multicolumn{1}{c}{$\bolds{\gamma=0}$}
 &\multicolumn{1}{c}{$\bolds{\mu_y = 0}$}&\multicolumn{1}{c}{$\bolds{\sigma_y = 1}$}
\\
\hline
Gaussian ML&50 & 0.0000&0.0054&-0.0175&0.0000&0.9943&0.7053\\
&100 & 0.0000&0.0016&-0.0084&0.0000&0.9812&0.7410\\
& 250
&0.0000&-0.0029&-0.0009&0.0000&0.9997&0.6917\\
&1\mbox{,}000 &0.0000&0.0005&-0.0013&0.0000&0.9788&0.7105\\
[3pt]
 {IGMM}& 50 &-0.0015&0.0054&-0.0060&0.4567&0.9945&0.7059\\
& 100&-0.0012&0.0015&-0.0030&0.4368&0.9813&0.7405\\
 & 250 &
0.0001&-0.0017&-0.0009&0.4210&0.9997&0.6919\\
& 1\mbox{,}000&0.0003&0.0005&-0.0008&0.4014&0.9788&0.7102\\
[3pt]
Lambert $W$ ML&50 &-0.0013&0.0054&-0.0126&0.5144&0.9951&0.7210\\
& 100&-0.0013&0.0016&-0.0072&0.4670&0.9813&0.7407\\
 & 250
&0.0002&-0.0017&-0.0027&0.4333&0.9997&0.6922\\
&1\mbox{,}000&0.0003&0.0005&-0.0012&0.4039&0.9788&0.7106\\
[3pt]
Skew-normal ML&50 & \multicolumn{1}{c}{\textit{NA}}&0.0052&-0.0135&\multicolumn{1}{c}{\textit{NA}}&0.9928&0.7149\\
& 100& \multicolumn{1}{c}{\textit{NA}}&0.0015&-0.0073&\multicolumn{1}{c}{\textit{NA}}&0.9821& 0.7409\\
 & 250 & \multicolumn{1}{c}{\textit{NA}}&-0.0018&-0.0027&\multicolumn{1}{c}{\textit{NA}}&1.0004&
0.6925\\
&1\mbox{,}000& \multicolumn{1}{c}{\textit{NA}}&0.0000&-0.0013&\multicolumn{1}{c}{\textit{NA}}&0.9788& 0.7105\\
\hline
\end{tabular*}
\end{table}

Table \ref{tab:N01_sim} displays the bias and root mean square error
({RMSE}) of $\widehat{\theta}$. Not only are all estimators
unbiased, but they also have essentially equal  {RMSE} for
$\widehat{\mu}_y$ and $\widehat{\sigma}_y$. It is well known that
the Gaussian MLE of $\sigma_x$ is only asymptotically unbiased, but
for small samples it underestimates the standard deviation, whereas a
method of moments estimator such as IGMM does not have that problem
(see $N = 50$). For $\widehat{\gamma}$ the  {IGMM} estimator
has slightly smaller  {RMSE} than  {MLE} for small $N$;
for large $N$ the difference disappears. This can also be explained by
an only asymptotically unbiased MLE for $\sigma_x$, and the functional
relation \eqref{eq:sigma_y_gamma_sigma_x} of $\gamma$, $\sigma_x$
and $\sigma_y$.

Overall, estimating $\gamma$ has \textit{no} effect on the quality of
the remaining parameter estimates, if the data comes from a truly
(symmetric) Gaussian distribution. A~Shapiro Wilk Gaussianity test on
the $n = 1\mbox{,}000$ estimates of $\widehat{\gamma}_{\mathrm{IGMM}}$ and $\widehat
{\gamma}_{\mathrm{MLE}}$ gives $p$-values of $68.91\%$ and $68.25\%$,
respectively ($N = 1\mbox{,}000$), and thus confirms the asymptotic normality
of $\widehat{\gamma}$ as stated in Section
\ref{sec:LambertW_MLE}.

\subsection{\texorpdfstring{Slightly skewed data: $\gamma= -0.05$}{Slightly skewed data: gamma = -0.05}}
This choice of $\gamma$ is motivated by real world data---in
particular, asset returns typically exhibit slightly negative skewness
[$\gamma_1(-0.05) = -0.30063$].

%
\begin{table}
\tabcolsep=0pt
\caption{Bias and  {RMSE} of
$\widehat{\theta}$ for $\gamma=-0.05$ and $X \sim N(\mu_x(\gamma
),\sigma_x^2(\gamma))$}
\label{tab:typical_LambertW}
\begin{tabular*}{\textwidth}{@{\extracolsep{\fill}}ld{5.0}d{2.4}d{2.4}d{2.4}d{1.4}d{1.4}d{1.4}@{}}
\hline
 &&\multicolumn{3}{c}{\textbf{Bias}}&\multicolumn{3}{c@{}}{\textbf{RMSE}${}\bolds{\cdot}{}\bolds{\sqrt{N}}$}\\[-5pt]
&&\multicolumn{3}{c}{\hrulefill}&\multicolumn{3}{c@{}}{\hrulefill}\\
 &
\multicolumn{1}{c}{$\bolds N$}&\multicolumn{1}{c}{$\bolds{\gamma= -0.05}$}&\multicolumn{1}{c}{$\bolds{\mu_y = 0}$}
&\multicolumn{1}{c}{$\bolds{\sigma_y = 1}$} &\multicolumn{1}{c}{$\bolds{\gamma= -0.05}$}
&\multicolumn{1}{c}{$\bolds{\mu_y = 0}$}&\multicolumn{1}{c@{}}{$\bolds{\sigma_y = 1}$}
\\
\hline
Gaussian ML&50 &  0.0500&0.0057&-0.0176&0.3536&0.9952& 0.7350 \\
&100 &  0.0500&0.0016&-0.0083&0.5000&0.9826&0.7741 \\
& 250 &
0.0500&-0.0029&-0.0009&0.7906&0.9981&0.7095 \\
&1\mbox{,}000 & 0.0500&0.0005&-0.0014&1.5811&0.9781&0.7281 \\
[3pt]
 {IGMM}& 50 & -0.0008&0.0046&-0.0057&0.4582&0.9954&0.7410\\
& 100& -0.0008&0.0011&-0.0026&0.4389&0.9828&0.7753 \\
 & 250 &
0.0002&-0.0019&-0.0007&0.4189&0.9982&0.7102\\
& 1\mbox{,}000& 0.0003&0.0005&-0.0009&0.3986&0.9780&0.7276\\
[3pt]
Lambert $W$ ML& 50 & -0.0043&0.0052&-0.0116&0.5113&0.9961&0.7570\\
& 100& -0.0029&0.0015&-0.0062&0.4701&0.9829&0.7802\\
 & 250 &
-0.0006&-0.0017&-0.0024&0.4282&0.9981&0.7114\\
&1\mbox{,}000& 0.0001&0.0005&-0.0013&0.3992&0.9781&0.7284\\
[3pt]
Skew-normal ML&50 & \multicolumn{1}{c}{\textit{NA}}&0.0073&-0.0136&\multicolumn{1}{c}{\textit{NA}}&1.0011&0.7490\\
& 100& \multicolumn{1}{c}{\textit{NA}}&0.0026&-0.0067&\multicolumn{1}{c}{\textit{NA}}&0.9834& 0.7811\\
 & 250 & \multicolumn{1}{c}{\textit{NA}}&-0.0014&-0.0025&\multicolumn{1}{c}{\textit{NA}}&0.9990&
0.7109 \\
&1\mbox{,}000& \multicolumn{1}{c}{\textit{NA}}&0.0000&-0.0012&\multicolumn{1}{c}{\textit{NA}}&0.9796& 0.7281\\
\hline
\end{tabular*}
\end{table}

Table \ref{tab:typical_LambertW} presents the effect of ignoring small
asymmetry in data. Gaussian  {MLE} is by definition biased for
$\gamma$, but $\widehat{\mu}_y$ and $\widehat{\sigma}_y$ are still
good estimates. Neither  {IGMM} nor Lambert $W$  {MLE}
gives biased $\widehat{\theta}$, but the  {RMSE} of~$\widehat
{\sigma}_y$ increases for all estimators and all sample sizes. Again
 {IGMM} presents smaller  {RMSE} for $\widehat{\gamma}$
than  {MLE} for small $N$, but not for large $N$---for the same
reason as in the $\gamma= 0$ case. Notably, the skew-normal MLE for
$\mu_y$ and~$\sigma_y$ is also unbiased and has the same RMSE as the
Lambert $W$ and Gaussian competitors, even though the true distribution
is a Lambert $W{}\times{}$Gaussian, not a skew-normal.

\subsection{\texorpdfstring{Extremely skewed data: $\gamma= 0.3$}{Extremely skewed data: gamma = 0.3}}
In this case, the Lambert $W$ MLE should work better than the skew-normal
MLE, since the skewness coefficient $\gamma_1(0.3) = 1.9397$ lies
outside the theoretically possible values of skew-normal distributions.
Furthermore, the nonprincipal branch of the Lambert $W$ function becomes
more important as $p_{-1} \approx4.29 \cdot10^{-4}$, so the Lambert $W$
 {MLE} should also outperform  {IGMM}, which ignores the
nonprincipal solution.

Only the skew-normal MLE fails to provide accurate estimates of
location and scale for heavily skewed data sets; all other estimators
are practically unbiased (Table \ref{tab:extreme_LambertW}). The
 {RMSE} for $\widehat{\sigma}_y$ almost doubled compared to
the symmetric case, and for Gaussian as well as skew-normal MLE it is
increasing with sample size instead of decreasing. While $\widehat
{\gamma}_{\mathrm{IGMM}}$ has less bias, $\widehat{\gamma}_{\mathrm{MLE}}$ has a much
smaller  {RMSE}: not ignoring the nonprincipal branch more than
compensates the finite sample bias in $\widehat{\sigma}_x$.
Surprisingly, the  {RMSE} for $\widehat{\gamma}$ has
diminished by about $35\%$ over all sample sizes compared to the
symmetric case.

%
\begin{table}
\tabcolsep=0pt
\caption{Bias and  {RMSE} of
$\widehat{\theta}$ for $\gamma=0.3$ and $X \sim N(\mu_x(\gamma
),\sigma_x^2(\gamma))$}\label{tab:extreme_LambertW}
\begin{tabular*}{\textwidth}{@{\extracolsep{\fill}}ld{5.0}d{2.4}d{2.4}d{2.4}d{1.4}d{1.4}d{1.4}@{}}
\hline
 &&\multicolumn{3}{c}{\textbf{Bias}}&\multicolumn{3}{c@{}}{\textbf{RMSE}${}\bolds{\cdot}{}\bolds{\sqrt{N}}$}\\[-5pt]
&&\multicolumn{3}{c}{\hrulefill}&\multicolumn{3}{c@{}}{\hrulefill}\\
&\multicolumn{1}{c}{$\bolds{N}$}&\multicolumn{1}{c}{$\bolds{\gamma=0.3}$}&\multicolumn{1}{c}{$\bolds{\mu_y = 0}$}&\multicolumn{1}{c}{$\bolds{\sigma_y = 1}$} &\multicolumn{1}{c}{$\bolds{\gamma=0.3}$}&\multicolumn{1}{c}{$\bolds{\mu_y = 0}$}&\multicolumn{1}{c@{}}{$\bolds{\sigma_y = 1}$}
\\ \hline
Gaussian ML&50 &-0.3000&0.0029&-0.0336&2.1213&0.9851&1.2941\\
&100 &-0.3000&0.0006&-0.0194&3.0000&0.9863&1.3957 \\
& 250 & -0.3000&
-0.0076&-0.0056&4.7434&1.0045&1.4499\\
&1\mbox{,}000 & -0.3000& 0.0002&-0.0013&9.4868&0.9883&1.4954\\
[3pt]
 {IGMM}& 50 &-0.0076&0.0081&-0.0057&0.4374&0.9917&1.2417\\
&100& -0.0055&0.0028&-0.0063&0.4005&0.9853&1.2440 \\
 & 250
&-0.0032&-0.0012&-0.0056&0.3647&1.0009&1.2204 \\
&1\mbox{,}000&-0.0026&-0.0003&-0.0049&0.3197&0.9820&1.1992\\
[3pt]
Lambert $W$ ML&50 & 0.0180&0.0221& 0.0266&0.3844&1.0152&1.2385\\
& 100&  0.0115&0.0131& 0.0168&0.3241&1.0095&1.2218\\
 & 250 &  0.0055& 0.0053&
0.0054&0.2747&1.0102&1.1535\\
&1\mbox{,}000& 0.0000& 0.0021&-0.0021&0.2349&0.9818&1.1383\\
[3pt]
Skew-normal ML&50 & \multicolumn{1}{c}{\textit{NA}}&0.0695&-0.0938&\multicolumn{1}{c}{\textit{NA}}&1.3638&1.1965\\
& 100& \multicolumn{1}{c}{\textit{NA}}&0.0558&-0.0834&\multicolumn{1}{c}{\textit{NA}}&1.3508& 1.3182\\
 & 250 & \multicolumn{1}{c}{\textit{NA}}& 0.0520&-0.0748&\multicolumn{1}{c}{\textit{NA}}&1.4865&
1.5577\\
&1\mbox{,}000& \multicolumn{1}{c}{\textit{NA}}& 0.0560&-0.0704&\multicolumn{1}{c}{\textit{NA}}&2.1588& 2.4585\\
\hline
\end{tabular*}
\vspace*{-4pt}
\end{table}

\subsubsection*{Discussion}
Estimation of $\mu_y$ is unaffected by the value of $\gamma$; the
quality of $\widehat{\sigma}_y$, however, depends on $\gamma$: the
larger $\gamma$, the greater the  {RMSE} of~$\widehat{\sigma}_y$.
For $\gamma= 0$ the Lambert $W$ methods perform equally well as
Gaussian  {MLE}, whereas for nonzero $\gamma$ Gaussian and---to
some extent---skew-normal  {MLE} have inferior qualities
compared to the Lambert $W$ alternatives. In particular, the RMSE for
$\widehat{\sigma}_y$ increases with sample size.

Hence, there is no gain restricting analysis to the (symmetric)
Gaussian case, as the Lambert $W$ framework extends this distribution to
a broader class, without losing the good properties of Gaussian MLE.
For little asymmetry in the data, both the Lambert $W$ and the
skew-normal approach give accurate and precise estimates of location,
scale and skewness. Yet for heavily skewed data (skewness greater than
$0.995$ in absolute value), the skew-normal framework fails not only in
theory, but also in practice to provide a good approximation.

%
\begin{table}
\tabcolsep=0pt
\caption{Average number of iterations
($\mathit{tol} = 10^{-6}$): (top) IGMM Algorithm \protect\ref{alg:IGMM} including the
iterations in Algorithm \protect\ref{alg:gamma_GMM}; (bottom) IGMM only (not
counting iterations in Algorithm \protect\ref{alg:gamma_GMM}). (left) Gaussian
input; (right) student-$t$ input with $\nu= 4$ degrees of freedom. Based
on $n = 1\mbox{,}000$ replications}\label{tab:count_iterations}
\begin{tabular*}{\textwidth}{@{\extracolsep{\fill}}lcd{2.2}d{2.2}d{2.2}d{2.2}d{2.2}@{}}
\hline
&\multicolumn{6}{c@{}}{$\bolds\gamma$}\\[-5pt]
&\multicolumn{6}{c@{}}{\hrulefill}\\
 $\bolds N$   & \multicolumn{1}{c}{$\bolds 0$} & \multicolumn{1}{c}{$\bolds{-0.05}$} & \multicolumn{1}{c}{$\bolds{0.3}$} & \multicolumn{1}{c}{$\bolds{0, \nu= 4}$} &
\multicolumn{1}{c}{$\bolds{-0.05, \nu= 4}$} & \multicolumn{1}{c@{}}{$\bolds{0.3, \nu= 4}$}\\
\hline
\hphantom{10\mbox{,}}$50$ & 8.39 & 10.24 & 34.65 & 15.82 & 16.76 & 26.91 \\
\hphantom{1\mbox{,}}$100$ & 6.05 & 8.16 & 35.01 & 17.52 & 19.12 & 21.45\\
\hphantom{1\mbox{,}}$250$ & 4.37 & 6.45 & 27.51 & 17.83 & 21.34 & 13.23 \\
$1\mbox{,}000$ & 3.58 & 4.96 & 18.43 & 17.20 & 24.58 & 6.49 \\
 [3pt]
\hphantom{10\mbox{,}}$50$ & 4.43 & 4.60 & 6.24 & 4.56 & 4.64 & 6.23\\
\hphantom{1\mbox{,}}$100$ & 4.10 & 4.44 & 6.44 & 4.46 & 4.44 & 5.78 \\
\hphantom{1\mbox{,}}$250$ & 3.90 & 4.26 & 5.92 & 4.15 & 4.19 & 5.45 \\
$1\mbox{,}000$ & 3.58 & 4.11 & 5.91 & 3.78 & 4.04 & 5.36 \\
\hline
\end{tabular*}
\end{table}

Table \ref{tab:count_iterations} shows the average number of
iterations the IGMM algorithm needed to converge: for increasing sample
size it needs less iterations---sample moments can be estimated more
accurately; more iterations are needed for larger $\gamma$---as the
starting value for $\gamma$ is based on the Taylor expansion around
$\gamma=0$ and moving away from the origin makes the initial estimate
$\gamma^{(0)} := \widehat{\gamma}_{\,\mathrm{Taylor}}$ less precise.

A closer look at the two sub-tables (top and bottom) shows that finding
the optimal $\gamma$ (Algorithm \ref{alg:gamma_GMM}) becomes much
more difficult\vadjust{\eject} for increasing $\gamma$ and sample size $N$ than
finding the optimal $\mu_x$ and $\sigma_x$ given the optimal~$\widehat{\gamma}_{\mathrm{GMM}}$
(Algorithm \ref{alg:IGMM}). For $\gamma=
0$ and large $N$ there is almost no difference between the total number
of iterations (top) and the number of iterations in Algorithm \ref
{alg:IGMM} only (bottom). For large $\gamma$, however, the total
number of iterations is approximately $5$ times as large. The right
panel shows the values for simulations of a Lambert $W\times t$ RV
with $\nu= 4$ degrees of freedom. For small~$\gamma$, finding
$\widehat{\gamma}_{\mathrm{GMM}}$ takes much longer than for Gaussian input;
surprisingly, for large $\gamma$ convergence is reached faster. This
is probably a result of the constrained optimization: due to more
extreme values for a $t$-distribution, Algorithm \ref{alg:gamma_GMM}
often returns one of the two boundary values for $\widehat{\gamma
}_{\mathrm{GMM}}$ without even starting the optimization process.

Given its good empirical properties, fairly general assumptions about
the input variable $X$, and its fast computation time, the
{IGMM} algorithm can be used as a quick Lambert $W$ check. For a
particular Lambert $W\times F$ distribution, the Lambert $W \times
 F$  {MLE} gives more accurate results, especially for heavily
skewed data.

\section{Applications}
\label{sec:applications}

This section demonstrates the usefulness of the presented methodology
on real world data. In the first example I analyze parts of the
Australian Athletes data set\footnote{R package \texttt{LambertW},
data set \texttt{AA}.} which can be typically found in the literature
on modeling skewed data [\citet{Genton05}, \citet{AzzaliniDallaValle96}].
%
%
\begin{figure}

\includegraphics{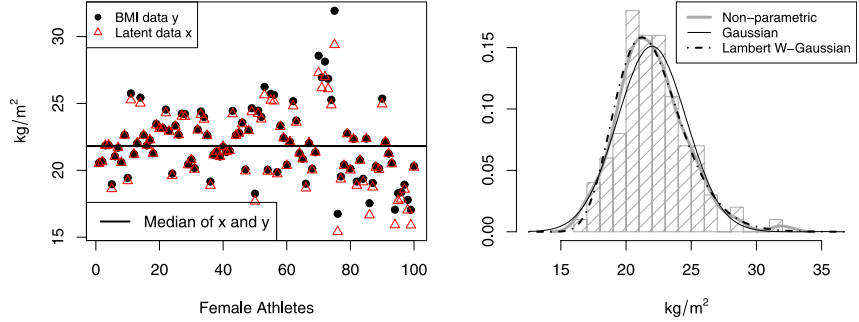}

\caption{Australian Athletes BMI:  {(left)} observed
data $\mathbf{y}$ (dots) and back-transformed
data~$\widehat{\mathbf{x}}_{\widehat{\tau}_{\mathrm{MLE}}}$ (triangles);  {(right)}
histogram plus density estimates.}\label{fig:BMI}
\vspace*{4pt}
\end{figure}

%
\begin{table}[b]
\vspace*{4pt}
\tabcolsep=0pt
\tablewidth=220pt
\caption{BMI ($\mathbf{y}$) and
back-transformed data $\widehat{\mathbf{x}}_{\widehat{\tau
}_{\mathrm{IGMM}}}$ and~$\widehat{\mathbf{x}}_{\widehat{\tau
}_{\mathrm{MLE}}}$: (top) summary statistics; (bottom) Shapiro--Wilk
(SW), Jarque--Bera~(JB)~normality tests}\label{tab:summary_BMI}
\begin{tabular*}{220pt}{@{\extracolsep{\fill}}ld{2.3}d{2.3}d{2.3}@{}}
\hline
 \textbf{BMI} & \multicolumn{1}{c}{$\mathbf{y}$} &  \multicolumn{1}{c}{$\bolds{\widehat{\mathbf{x}}_{\widehat{\tau}_{\mathrm{IGMM}}}}$} & \multicolumn{1}{c@{}}{$\bolds{\widehat{\mathbf{x}}_{\tau(\widehat{\theta}_{\mathrm{MLE}})}}$}
\\
\hline
Min & 16.750 & 15.356 & 15.406 \\
Max & 31.930 & 29.335 & 29.384 \\
Mean & 21.989 & 21.735 & 21.742 \\
Median & 21.815 & 21.815 & 21.815 \\
St. dev. & 2.640 & 2.570 & 2.569 \\
Skewness & 0.683 & 0.000 & 0.017 \\
Kurtosis & 1.093 & 0.186 & 0.187 \\
[6pt]
SW & 0.035 & 0.958 & 0.959 \\
JB & 0.001 & 0.877 & 0.874\\
\hline
\end{tabular*}
\end{table}

The second example reexamines the LATAM returns introduced in Section
\ref{sec:introduction}. A Lambert $W \times t$-distribution is found
to give an appropriate fit, both for the raw data as well as the
standardized residuals of an auto-regressive conditional
heteroskedastic time series model (see Section \ref{sec:non_iid_GARCH}
for details). In particular, a comparison of risk estimators (Value at
Risk) demonstrates the suitability of the Lambert $W\times F$
distributions to model financial data.\looseness=1

\subsection{BMI of Australian athletes}
Figure \ref{fig:BMI} shows the Body Mass Index (BMI) of $100$ female
Australian athletes (dots) and Table \ref{tab:summary_BMI} lists
several statistical properties (column 1). Although the data appear
fairly Gaussian, its large positive skewness makes both tests reject
normality on a $5\%$ level.\looseness=1

After 5 iterations $\widehat{\tau}_{\mathrm{IGMM}} = (21.735, 2.570,
0.099)$, which implies $\widehat{\mu}_y =\break 21.992$, $\widehat{\sigma
}_y = 2.633$, and $\gamma_1(\widehat{\gamma}_{\mathrm{IGMM}})=0.601$,
assuming Gaussian input.

\begin{table}
\caption{Lambert $W{}\times{}$Gaussian
{MLE} for the BMI data}\label{tab:MLE_BMI}
\begin{tabular}{@{}ld{2.3}cd{2.3}c@{}}
\hline
 &\multicolumn{1}{c}{\textbf{Estimate}}&\multicolumn{1}{c}{\textbf{Std. error}}&\multicolumn{1}{c}{$\bolds t$ \textbf{value}}&\multicolumn{1}{c@{}}{$\bolds{\operatorname{Pr}(> \mid t \mid)}$}
\\
\hline
$\mu_x$ & 21.742 & 0.274 & 79.494 & 0.000 \\
$\sigma_x$ & 2.556 & 0.188 & 13.618 & 0.000 \\
 [3pt]
$\gamma$ & 0.096 & 0.039 & 2.481 & 0.013 \\
\hline
\end{tabular}
\end{table}

The BMI data set consists of exactly $n=100$ i.i.d. samples and Table
\ref{tab:N01_sim} lists finite sample properties of $\widehat{\gamma
}_{\mathrm{IGMM}}$ for this case.\footnote{Although $\mathbf{y}$ is clearly
not $\mathcal{N}(0,1)$, the location-scale invariance of Lambert
$W{}\times{}$Gaussian RVs makes this difference to scenario 1 in the
simulations [$Y \equiv X \sim\mathcal{N}(0,1)$] irrelevant; finite
sample properties of $\gamma$ do not change between $X \sim\mathcal
{N}(0,1)$ and general $X \sim\mathcal{N}(\mu_x, \sigma_x^2)$, since
in both cases $\mu_x$ and $\sigma_x$ are also estimated.} Thus, if $Y
= \mathit{BMI}$ was Gaussian, then
%
%
\begin{equation}
\label{eq:BMI_gamma_finite_sample_distribution}
\frac{\sqrt{100} \widehat{\gamma}_{\mathrm{IGMM}}}{0.4368} \sim
\mathcal{N}(0,1).
\end{equation}

Plugging $\widehat{\gamma}_{\mathrm{IGMM}} = 0.099$ into \eqref
{eq:BMI_gamma_finite_sample_distribution} gives $2.279$ and a
corresponding $p$-va\-lue of $0.0113$. Thus, $\widehat{\gamma}_{\mathrm
{IGMM}}$ is significant on a $5\%$ level, yielding an indeed positively
skewed distribution for the  {BMI} data $\mathbf{y}$.

As both tests cannot reject Gaussianity for $\widehat{\mathbf
{x}}_{\widehat{\tau}_{\mathrm{IGMM}}}$, a Lambert $W{}\times
{}$Gaussian approach seems reasonable. Table \ref{tab:MLE_BMI} shows that
all estimates are highly significant, where standard errors are
obtained by numerical evaluation of the Hessian at the optimum. As not
one single test can reject normality of $\widehat{\mathbf
{x}}_{\widehat{\tau}_{\mathrm{MLE}}}$ (triangles in Figure~\ref
{fig:BMI}), an adequate model to capture the statistical properties of
the BMI data is
\begin{eqnarray*}
\mathit{BMI} =  ( U e^{0.099 U}  ) 2.556 + 21.742, \qquad U = \frac
{X-21.742}{2.556} \sim\mathcal{N}(0,1).
\end{eqnarray*}

For $\widehat{\theta}_{\mathrm{MLE}}$ the support of $\mathit{BMI}$ lies in
the half-open interval $[11.967, \infty)$.
As all observations lie within these boundaries, $\widehat{\theta
}_{\mathrm{MLE}}$ is indeed a (local) maximum. Figure \ref{fig:BMI}
shows the closeness of the Lambert $W{}\times{}$Gaussian density to the
histogram and kernel density estimate, whereas the best Gaussian is
apparently an improper approximation.

Although a more detailed study of athlete type and other health
indicators might explain the prevalent skewness, the Lambert $W$ results
at least support common sense: the human body has a natural
physiological lower bound\footnote{The lower truncation of the BMI at
$11.967$ corresponds to a $180$~cm tall athlete only weighing $38.88$
kg.} for the BMI, whereas outliers on the right tail---albeit, in
principle, also having an upper bound---are more likely.

%
\subsection{Asset returns}
\label{sec:financial_data}
A lot of financial data, also the LATAM return series introduced in
Section \ref{sec:introduction} (Table \ref{tab:LATAM_summary} and
Figure \ref{fig:LATAM_news4}), display negative skewness and excess
kurtosis. These so-called \textit{stylized facts} are well known and
typically addressed via (generalized) auto-regressive conditional
heteroskedastic (GARCH) [\citet{Engle82}, \citet{Bollerslev86}] or stochastic
volatility (SV) models [\citet{MelinoTurnbull90}, \citet{DeoHurvichYu06}].
A theoretical analysis of Lambert $W\times F$ time series models,
however, is far beyond the scope and focus of this work. For empirical
evidence regarding the usefulness and significance of Lambert
$W\times F$ distributions in GARCH models and possible future research
directions see Section \ref{sec:non_iid_GARCH}. It is also worth
noting that the Lambert $ W\times F$ transformation \eqref
{eq:noncentral_nonscale_LambertW_Y} resembles SV models very closely,
and connections between the two can be made in future work.

Based on the news $\leftrightarrow$ return interpretation in a stock
market $\mathcal{S}$, it makes sense to assume a symmetric input
distribution $F_X(x)$ for the latent news RV $X$. Without specifying
the symmetric $F_X(x)$ any further, the  {IGMM} algorithm gives
a robust estimate for $\tau$: here $\widehat{\tau}_{\mathrm{IGMM}}
= (-0.048, 0.190, 1.456)$. 
Column 2 of Table \ref{tab:LATAM_summary} shows that the unskewed data
$\widehat{\mathbf{x}}_{\widehat{\tau}_{\mathrm{IGMM}}}$---here
interpreted as news hitting the market---is non-Gaussian, but a
$t$-distribution cannot be rejected. In consequence, $Y$ is modeled as a
Lambert $W{}\times{}$location-scale $t$-distribution with $\bolds
\beta= (c,s, \nu)$, where $c$ is the location, $s$ the scale and $\nu
$ the degrees of freedom parameter.\vadjust{\goodbreak} Table \ref{tab:LATAM_lambertW_t}
shows that all coefficients of $\widehat{\theta}_{\mathrm{MLE}}$ are
highly significant; in particular, $\widehat{\gamma}$ increased
substantially (in absolute value), as $\gamma$ now solely addresses
asymmetry in the data, and $\nu$ can capture excess kurtosis. Thus,
the prevalent negative skewness in the LATAM daily returns is not an
artifact of large outliers in the left tail of an otherwise symmetric
distribution, but a significant characteristic of the data.

%
\begin{table}
\caption{Lambert $W\times t$ MLE for
the  {LATAM} series}\label{tab:LATAM_lambertW_t}
\begin{tabular}{@{}ld{2.3}cd{2.3}c@{}}
\hline
& \multicolumn{1}{c}{\textbf{Estimate}} &\multicolumn{1}{c}{\textbf{Std. error}} & \multicolumn{1}{c}{$\bolds t$ \textbf{value}} & $\bolds{\operatorname{Pr}(> \mid t \mid)}$ \\
\hline
$c$ & 0.197 & 0.037 & 5.270 & 0.000 \\
$s$ & 1.240 & 0.057 & 21.854 & 0.000 \\
$\nu$ & 7.047 & 2.196 & 3.208 & 0.001 \\
 [3pt]
$\gamma$ & -0.053 & 0.014 & -3.860 & 0.000 \\
\hline
\end{tabular}
\end{table}

In order to check if the Lambert $W\times t$-distribution is indeed
an appropriate model for $\mathbf{y}$, it is useful to study the
back-transformed data $\widehat{\mathbf{x}}_{\widehat{\tau}}$; here
$\widehat{\tau}_{\mathrm{MLE}} := \tau(\widehat{\theta}_{\mathrm
{MLE}}) = (0.197, 1.465, -0.053)$. Not surprisingly, the skewness
of~$\widehat{\mathbf{x}}_{\widehat{\tau}_{\mathrm{MLE}}}$ reduced to
almost $0$ (column 3 of Table \ref{tab:LATAM_summary}). As a
Kolmogorov--Smirnov test cannot reject a student $t$-distribution for
$\widehat{\mathbf{x}}_{\widehat{\tau}_{\mathrm{MLE}}}$, the
Lambert~$W{}\times{}t$-distribution
\begin{eqnarray}
Y &=&  ( U e^{-0.053 U}  ) 1.465 + 0.197, \nonumber\\
U &=& \frac{X - 0.197}{1.465}, \qquad U \sqrt{\frac
{7.047}{7.047-2}} \sim t_{\nu= 7.047} \nonumber
\end{eqnarray}
is an adequate unconditional probabilistic model for the
{LATAM} returns $\mathbf{y}$.

%
\begin{figure}

\includegraphics{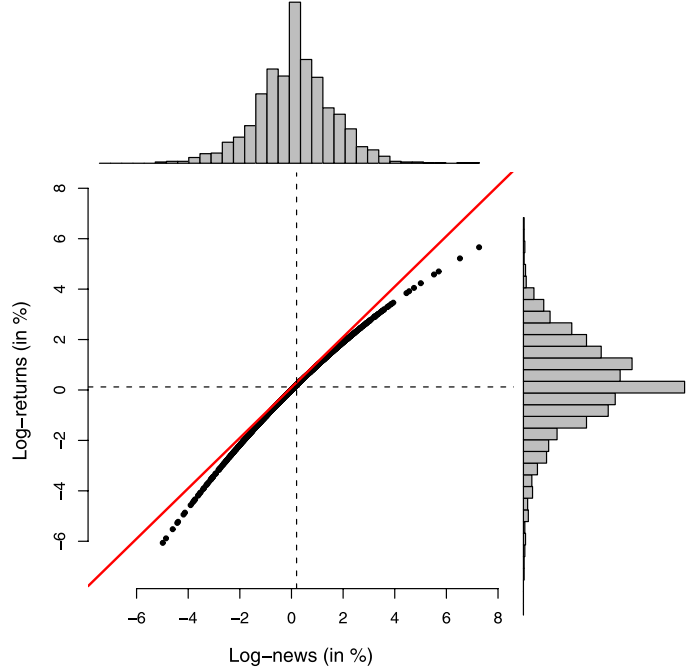}

\caption{News $\widehat{\mathbf
{x}}_{\widehat{\tau}_{\mathrm{MLE}}}$ $\leftrightarrow$ return
$\mathbf{y}$ scatter plot plus histograms; solid $45^{\circ}$ line:
$\gamma= 0$. Dashed vertical and horizontal lines represent the sample
mean of $\widehat{\mathbf{x}}_{\widehat{\tau}_{\mathrm{MLE}}}$ and
$\mathbf{y}$, respectively.}\label{fig:LATAM_news_return}
\end{figure}

The effect of news $x_t$ in the market $\mathcal{S}$ is clearly
shown in a scatter plot of $\widehat{\mathbf{x}}_{\widehat{\tau}_{\mathrm{MLE}}}$ versus $\mathbf{y}$.
For example,
consider the lower-left point $(x_{1346}, y_{1346}) \approx(-4.8,
-6.1)$ in Figure \ref{fig:LATAM_news_return}. Here, the observed
negative return equals $-6.1\%$, but as $\widehat{\gamma} = -0.053 <
0$, this outcome was an overreaction to bad news that was only
``worth'' $-4.8\%$. For location-scale Lambert $W$ RVs the skewness
parameter $\gamma$ is a powerful, yet easy way to characterize
different markets/assets. The negative $\widehat{\gamma}$ shows that
this specific market (system) is exaggerating bad news, and devalues
positive news.

\vspace*{-2pt}
\subsubsection*{Value at risk (VaR)} The \textsc{VaR} is a popular measure
in financial statistics to estimate the potential loss for an
investment in an asset over a fixed time period. That is, the maximum
percentage an investor can expect to lose---with a confidence of
$1-\alpha$---over a fixed time period. Statistically this corresponds
to the $\alpha$-quantile of the distribution. The \textsc{VaR} can be
obtained in various ways: the simplest are empirical and theoretical
quantiles given the estimated parameter vector of a parametric
distribution (which are sufficient for comparative
purposes).\vadjust{\eject}

%
\begin{table}
\caption{VaR comparison for the LATAM series}\label{tab:LATAM_VaR}
\begin{tabular}{@{}ld{2.3}d{2.3}d{2.3}d{1.3}d{1.3}d{1.3}d{1.3}@{}}
\hline
&\multicolumn{7}{c@{}}{$\bolds\alpha$}\\[-5pt]
&\multicolumn{7}{c@{}}{\hrulefill}\\
\textbf{Method}    & \multicolumn{1}{c}{\textbf{0.005}} & \multicolumn{1}{c}{\textbf{0.01}}
& \multicolumn{1}{c}{\textbf{0.05}} & \multicolumn{1}{c}{\textbf{0.5}} & \multicolumn{1}{c}{\textbf{0.95}} & \multicolumn{1}{c}{\textbf{0.99}} & \multicolumn{1}{c@{}}{\textbf{0.995}} \\
\hline
empirical & -4.562 & -4.078 & -2.478 & 0.138 & 2.344 & 3.192 & 3.818 \\
[3pt]
Gaussian & -3.660 & -3.294 & -2.293 & 0.121 & 2.535 & 3.535 & 3.901 \\
$t$ & -4.297 & -3.634 & -2.214 & 0.121 & 2.455 & 3.875 & 4.538 \\
Lambert $W\times t$ & -4.871 & -4.049 & -2.358 & 0.197 & 2.351 &
3.437 & 3.893 \\
Skew-$t$ & -4.715 & -3.973 & -2.364 & 0.201 & 2.346 & 3.465 & 3.957 \\
\hline
\end{tabular}
\end{table}

As expected, a Gaussian distribution underestimates both the low and
high quantiles, as it lacks the capability to capture excess kurtosis
(see Table \ref{tab:LATAM_VaR}). The $t$-distribution with $\widehat
{\nu}_{\mathrm{MLE}} = 6.22$ degrees of freedom has heavier tails, but
underestimates low and overestimates high quantiles: clearly an
indication of the prevalent skewness in the data. The Lambert
$W\times t$ and the skew $t$-distribution\footnote{MLE estimates
are $(0.917,
1.422,-0.799, 7.156)$ for the location, scale, shape and degrees of
freedom parameter respectively; function \texttt{st.mle} in the
\texttt{sn} package.} are the best approximation to the empirical
quantiles: both heavy tails and negative skewness are captured (see
also the Lambert $W\times t$ QQ plot in Figure \ref
{fig:LATAM_news4}). There is no clear ``winner'' between the two skewed
distributions: skew-$t$ quantiles are closer to the empirical ones for
small $\alpha$, Lambert $W\times t$ quantiles are closer for large
$\alpha$. Around the median ($\alpha= 0.5$) both skewed distributions
are far away from the true value: the reason being a high concentration
of close to $0$ returns in financial assets, so-called ``inliers'' [see
\citet{Breidt95}].

\subsubsection{Nonindependence of financial data}
\label{sec:non_iid_GARCH}
It is well known that financial return series $y_t$ typically exhibit
positive auto-correlation in their squares~$y_t^2$, which violates the
independence assumption of the MLE presented in Section~\ref
{sec:LambertW_MLE}. A standard parametric way to capture this
dependence is a GARCH model [\citet{Bollerslev86}, \citet{Engle82}], which
models the variance at time~$t$, $\sigma_t^2$, as a function of its
own past. A simple, yet very successful model for an uncorrelated $y_t$
is a $\operatorname{GARCH}(1,1)$,
\begin{eqnarray*}
y_t &=& \mu+ \varepsilon_t \sigma_t, \\
\sigma_t^2 & =& \omega+ \alpha y_{t-1}^2 + \beta\sigma_{t-1}^2,
\end{eqnarray*}
where $\varepsilon_t$ is a zero-mean, unit-variance i.i.d. sequence
[for technical details see \citet{Nelson90}, \citet{Engle82}]. Typically,
$\varepsilon_t \sim\mathcal{N}(0,1)$, but also student~$t$- or skew
$t$-distributions are used for more flexibility in the conditional
distribution of $\varepsilon_t$ given the information set $\Omega
_{t-1}$ available at time $t-1$ [\citet{Laurent02}]. \citet
{FrenchSchwertStambaugh87}\vadjust{\eject} also found that the standardized residuals
$(y_t - \widehat{\mu}) / \widehat{\sigma}_t$---which can be
considered as an i.i.d. sequence---still exhibit negative skewness
after fitting a Gaussian GARCH model to $S \& P$ $500$ returns.

After fitting a student-$t$ $\operatorname{GARCH}(1,1)$ model\footnote{Function \texttt
{garchFit($\cdot$)} in the \texttt{fGarch} package.} to the LATAM
return series $\mathbf{y}$, the Lambert $W\times t$ MLE fit for the
standardized residuals---which are approximately i.i.d. and thus do
not violate the MLE assumptions---still gives a highly significant
$\widehat{\gamma} = -0.048$ with a $p$-value of $0.000113$ (other
estimates are not shown here).

While I will not study Lambert $W{}\times{}$student-$t$ GARCH models in
detail, this example and the great flexibility of Lambert $W\times F$
distribution combined with the possibility to symmetrize skewed data
suggest that Lambert $W\times F$ GARCH (and SV) models are a
promising area of future research.

This analysis confirms previous findings that negative skewness is an
important feature of asset returns. For example, optimal portfolio
models based on skewed distributions lead to better suited decision
rules to react to asymmetric price movements. It also shows that
Lambert $W$ distributions model the characteristics of financial returns
as well as skew $t$-distributions, with the additional option to recover
symmetric latent data, which is not possible for RVs based on a
manipulation of the p.d.f. rather than a variable transformation.

\section{Relation to Tukey's $h$ distribution}
\label{sec:Tukeys_h}
During the final review process, Professor Andrew F. Siegel suggested
possible connections of Lambert $W$ distributions to Tukey's $g$--$h$
distribution [\citet{Tukey77}]
%
%
\begin{equation}
Z = \frac{ \exp(g U) - 1 }{g} \exp \biggl( \frac{h}{2} U^2
\biggr), \qquad h \geq0,
\end{equation}
where $U \sim\mathcal{N} ( 0, 1  )$. Here $g$ is the skew
parameter and $h$ controls the tail behavior of $Z$.

Although the underlying idea to introduce skewness is the same, the
specific transformations to get the skewness effects are different, and
so are the properties of the transformed RVs. 

For $g \rightarrow0$,
%
%
\begin{equation}
\label{eq:Tukey_h}
Z = U \exp \biggl( \frac{h}{2} U^2  \biggr)
\end{equation}
becomes symmetric. The RV $Z$ has Tukey's $h$ distribution and is
commonly used to model heavy-tails [\citet{Fischer06}, \citet{Field04}].
Equation \eqref{eq:Tukey_h} reveals a close link of Lambert $W \times
 F$ RVs to the existing statistics literature by noting that if $Z
\sim
h$, then $Z^2 = U^2 e^{h U^2}$ has a noncentral, nonscaled Lambert
$W{}\times{}\chi_1^2$ distribution with $\gamma= h$.

For further important connections between the Lambert $W$ function and
Tukey's $h$ distribution see \citet{Goerg10Gaussianize}.

%

\section{Discussion and outlook}
\label{sec:discussion_outlook}

Whereas the Lambert $W$ function plays an important role in mathematics,
physics, chemistry, biology and other fields, it has not yet been used
in statistics. Here I introduce it in an input/output setting to skew
and ``unskew'' RVs and data, respectively. 


Successful application to biomedical and financial data together with
the great flexibility with respect to the type of input RV $X$ of
Lambert $W\times F$ RVs promise a wide range of applications as
well as theoretical studies for particularly chosen input distributions.

Last but not least, a very pragmatic advantage of the
transformation-based Lambert $W\times F$ RVs compared to other
approaches to asymmetry: data can be ``unskewed'' using Lambert's $W$ function.

\section*{Acknowledgments}
I am grateful to Professor Wilfredo Palma for giving me the opportunity
to work at the Department of Statistics, Pontificia Universidad Cat\'{o}lica de Chile, Santiago, where I
completed important parts of this study.


Furthermore, I want to thank Professor Reinaldo Arellano-Valle,
Professor Cosma Shalizi, the Editor Professor Stephen Fienberg and two
anonymous referees for helpful comments and suggestions on the manuscript.


%

\printaddresses

\end{document}